\documentclass[acmsmall]{acmart}
\AtBeginDocument{%
  \providecommand\BibTeX{{%
    \normalfont B\kern-0.5em{\scshape i\kern-0.25em b}\kern-0.8em\TeX}}}

\setcopyright{acmlicensed}
\copyrightyear{XXX}
\acmYear{XXX}
\acmDOI{XXXXXXX.XXXXXXX}

\acmJournal{JACM}
\acmVolume{37}
\acmNumber{4}
\acmArticle{111}
\acmMonth{8}

\usepackage{microtype}
\usepackage{multirow}
\usepackage{paralist}
\usepackage{soul}

\begin{document}

\title{Survey on the Evaluation of Generative Models in Music}

\author{Alexander Lerch}
\authornote{The first author conceptualized the paper and led the paper writing. The remaining co-authors contributed equally to the paper writing and are listed alphabetically.}
\orcid{0000-0001-6319-578X}
\email{alexander.lerch@gatech.edu}

\affiliation{%
  \institution{Georgia Institute of Technology}
  \city{Atlanta}
  \country{USA}
  }

\author{Claire Arthur}
\authornotemark[1]
\email{claire.arthur@gatech.edu}
\orcid{0000-0002-5454-8384}

\affiliation{%
  \institution{Georgia Institute of Technology}
  \city{Atlanta}
  \country{USA}
  }

\author{Nick Bryan-Kinns}
\authornotemark[1]
\email{n.bryankinns@arts.ac.uk}
\orcid{0000-0002-1382-2914}

\affiliation{%
  \institution{University of the Arts London}
  \city{London}
  \country{UK}
  }

\author{Corey Ford} 
\authornotemark[1]
\email{c.j.ford@qmul.ac.uk}
\orcid{0000-0002-6895-2441}

\affiliation{%
  \institution{University of the Arts London}
  \city{London}
  \country{UK}
  }

\author{Qianyi Sun}
\authornotemark[1]
\email{qsun75@gatech.edu}
\orcid{0009-0002-9889-5617}

\affiliation{%
  \institution{Georgia Institute of Technology}
  \city{Atlanta}
  \country{USA}
  }

\author{Ashvala Vinay}
\authornotemark[1]
\email{ashvala@gatech.edu}
\orcid{0000-0002-2487-2052}

\affiliation{%
  \institution{NoneType Computing}
  \city{Atlanta}
  \country{USA}
  }


\def\redline{false}
\ifthenelse{\equal{\redline}{true}}
{
\newcommand{\strikethrough}[1]{\textcolor{red}{\st{#1}}}
\newcommand{\assignment}[1]{{\huge\textbf{[#1]}}}
\newcommand{\note}[1]{\textcolor{black}{#1}}

\newcommand{\alex}[1]{\textcolor{blue}{#1}}
\newcommand{\ash}[1]{\textcolor{blue}{#1}}
\newcommand{\rose}[1]{\textcolor{blue}{#1}}
\newcommand{\corey}[1]{\textcolor{blue}{#1}}
\newcommand{\nbk}[1]{\textcolor{blue}{#1}}
\newcommand{\claire}[1]{\textcolor{blue}{#1}}
}
{
\newcommand{\strikethrough}[1]{}
\newcommand{\assignment}[1]{{\huge\textbf{[#1]}}}
\newcommand{\note}[1]{\textcolor{black}{#1}}

\newcommand{\alex}[1]{\textcolor{black}{#1}}
\newcommand{\ash}[1]{\textcolor{black}{#1}}
\newcommand{\rose}[1]{\textcolor{black}{#1}}
\newcommand{\corey}[1]{\textcolor{black}{#1}}
\newcommand{\nbk}[1]{\textcolor{black}{#1}}
\newcommand{\claire}[1]{\textcolor{black}{#1}}
}
{}
\def\bNewTable{true}
\def\bWithRAI{false}

\begin{abstract}
    Research on generative systems in music has seen considerable attention and growth in recent years. A variety of attempts have been made to systematically evaluate such systems. 
    We present an interdisciplinary review of the common evaluation targets, methodologies, and metrics for the evaluation of both system output and model \strikethrough{usability} \corey{use}, covering subjective and objective approaches, qualitative and quantitative approaches, as well as empirical and computational methods. We examine the benefits and limitations of these approaches from a musicological, an engineering, and an HCI perspective.
    
\end{abstract}

\ccsdesc[500]{General and reference~Surveys and overviews}
\ccsdesc[500]{General and reference~Evaluation}
\ccsdesc[300]{General and reference~Metrics}
\ccsdesc[300]{Human-centered computing~HCI design and evaluation methods}

\keywords{Music, Evaluation, Generative AI, Survey}

\received{26 June 2024}
\received[revised]{8 August 2025}
\received[accepted]{12 September 2025}


\maketitle




\section{Introduction}\label{sec:intro}

In recent years, advances in system architecture and training methodologies of generative systems in machine learning have led to increasingly powerful systems that have been applied to a variety of tasks and use cases. Particularly prominent are systems in natural language processing (e.g., \cite{brown_language_2020}) and image models (e.g., \cite{karras_style-based_2019, ramesh_zero-shot_2021}).
Generative systems for music, while not featured as prominently in the media, have also seen considerable progress. \alex{Although} the concept of computer-generated music \alex{dates} back to the mid-20th century \cite{hiller_experimental_1959, supper_few_2001} and has been actively researched since then \cite{cope_algorithmic_2000, fernandez_ai_2013}, there has been a rapid increase in interest and publications in the past five years \cite{civit_systematic_2022}, mostly driven by advances in neural approaches. Inspired by \citet{herremans_functional_2017}, we categorize generative music systems based on
\begin{inparaenum}[(i)]
    \item architecture,
    \item output, and
    \item input or control.\footnote{References will be limited to a select number of representative systems.}
\end{inparaenum}
A meta-review by \citet{civit_systematic_2022} lists the prevalent \textit{architectures} with decreasing number of occurrences as Recurrent Neural Networks (RNNs) \cite{hadjeres_deepbach_2017, mao_deepj_2018}, Feed-Forward Networks (FF) \cite{yang_midinet_2017, dong_musegan_2018}, Variational Audio Encoders (VAEs) \cite{roberts_hierarchical_2018, dhariwal_jukebox_2020}, evolutionary algorithms \cite{stoltz_mu_psyc_2019, de_prisco_evocomposer_2020}, and Transformer-based approaches \cite{huang_music_2019, donahue_lakhnes_2019,copet_simple_2023}, sometimes combined with Generative Adversarial Networks (GANs) \cite{yang_midinet_2017, dong_musegan_2018}\ash{, with Diffusion networks \cite{evans_fast_2024, evans_stable_2025, liu_audioldm_2023, chen_musicldm_2024, schneider_mousai_2024, ning_diffrhythm_2025}, or with Rectified Flows \cite{li_omniflow_2025}}. Details of many typical neural architectures for music generation can be found in \citet{briot_deep_2020}. 

A key distinction of \textit{output} types of generative systems in music is whether the output is audio \cite{dieleman_challenge_2018, copet_simple_2023, agostinelli_musiclm_2023} or symbolic (e.g., MIDI, MusicXML, etc.) \cite{huang_music_2019, roberts_hierarchical_2018}. The length of the output can also vary: it might range from as short as a single note or event in the case of audio synthesizers \cite{engel_neural_2017, engel_ddsp_2020, kong_diffwave_2020} over phrases \cite{naruse_pop_2022, huang_pop_2020} to complete musical pieces \cite{dhariwal_jukebox_2020, lam_efficient_2023}. Furthermore, systems might generate a single-voiced melody \cite{yang_midinet_2017, hakimi_bebopnet_2020} or a polyphonic output with multiple voices \cite{hadjeres_deepbach_2017, herremans_morpheus_2019, oore_this_2020, huang_pop_2020}.

The \textit{input} of a system depends very much on the design goals. A system might require no input at all \cite{pasini_musika_2022}, a few parameters for conditioning \cite{chen_continuous_2020, eisenbeiser_latent_2020}, a text prompt \cite{copet_simple_2023, agostinelli_musiclm_2023}, a melody to be harmonized  \cite{hadjeres_deepbach_2017, yeh_automatic_2021}, or a musical phrase to be continued \cite{huang_music_2019, huang_pop_2020} or inpainted  \cite{pati_learning_2019,marafioti_gacela_2021, araneda-hernandez_musib_2023}.
While this variety of approaches and the multitude of available studies imply rapid progress, this progress is hard to quantify, and there is evidence that the quality improvements might not \alex{be} as dramatic as the number of publications suggests \cite{yin_deep_2023}.

All generative systems pose challenges in terms of evaluation since a ground truth target, or unique ``correct'' reference result, does not usually exist. Systems targeting the generation of artistic output are particularly difficult to assess due to the subjectivity of aesthetic assessment. 
The assessment of music poses a unique set of challenges due to 
\begin{inparaenum}[(i)]
    \item   its sequential yet highly structured form, 
    \item   the abstract musical language and the resulting unclear definition of content in music, 
    \item   the limited musical meaning of commonly-used music descriptors and the corresponding inadequacy to fully represent the multi-dimensionality of music and musical expression, 
    \item   the context-dependent interaction between expectation and surprise, and
    \item   the constant reinterpretation of musical ideas through music performance. 
\end{inparaenum}

These challenges have led to a large variety in approaches to system evaluation with a multitude of different evaluation targets, methodologies, and metrics. Inter-study inconsistencies in evaluation make \strikethrough{a}\alex{the} comparison of research results essentially impossible. If results cannot be compared, do not sufficiently reflect the actual quality of a system, or have been acquired in very different settings, the notion of progress in this field becomes questionable, as we cannot measure progress without relevant, commonly used metrics.
Despite these problems being recognized as important challenges \cite{yang_evaluation_2020, briot_deep_2020, liang_harmonizing_2023}, no general solutions have been proposed, and evaluation still seems to be largely neglected or treated as an afterthought. For instance, \citet{civit_systematic_2022} provide a meta-review of generative music systems but only mention evaluation in passing. \alex{\citet{zhao_ai-enabled_2025} review prompt-based generative music systems but refer only to the evaluation of creativity as an unsolved challenge.} While \citet{bandi_power_2023} present a dedicated evaluation section in their extensive review of generative systems, music is unfortunately not discussed. \alex{The only two exceptions are} \citet{ji_survey_2023} \alex{and \citet{wang_review_2024}, summarizing} some objective and subjective approaches to evaluation. 

\begin{figure}%
	\includegraphics[width=.8\columnwidth]{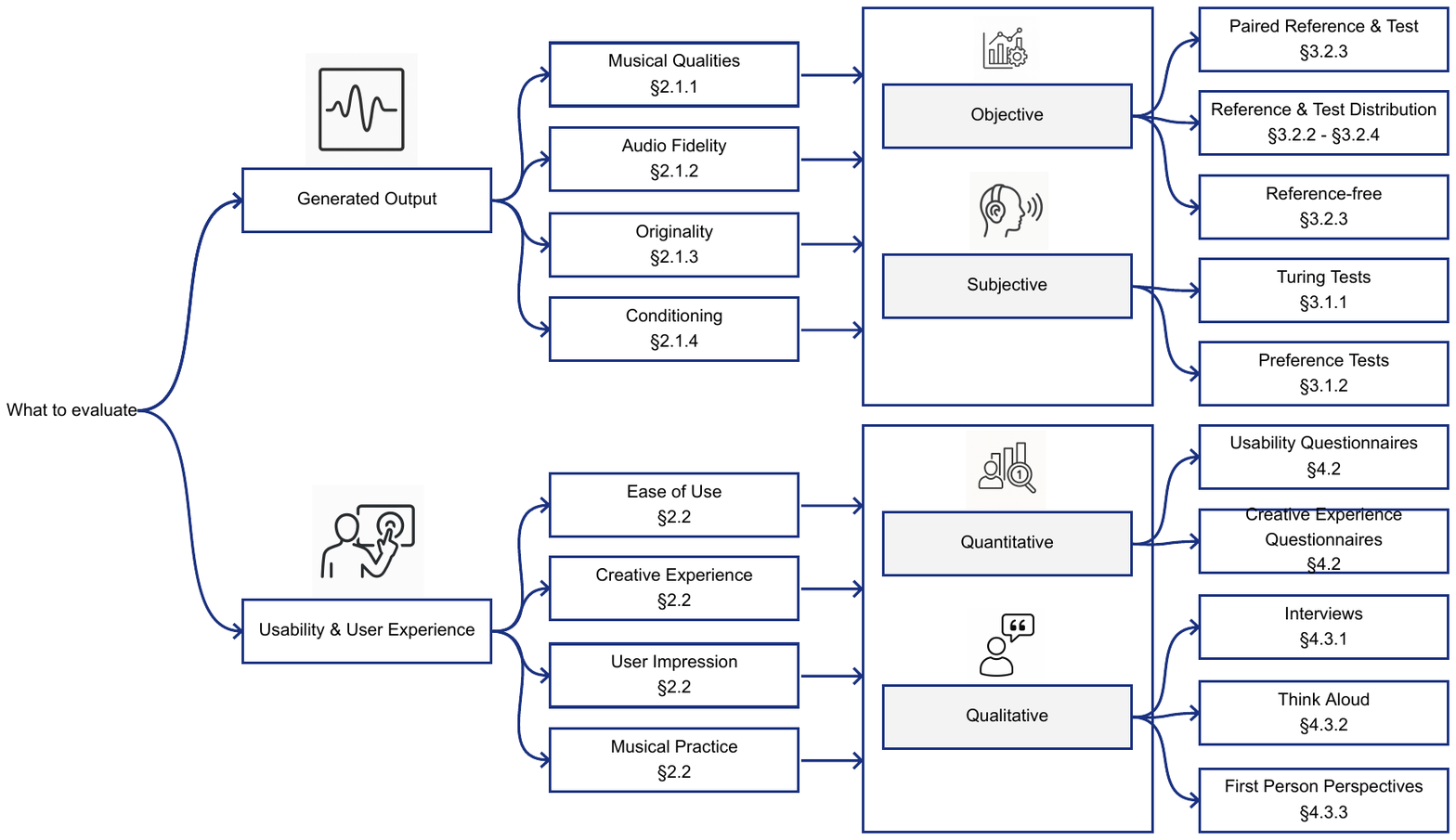}%
	\caption{\alex{Structure of the presented evaluation approaches and methodologies.}}%
  \Description{Structural flow-chart overview of the topics covered in the paper, starting with evaluation targets (musical qualities, audio fidelity, originality, and conditioning for output evaluation and ease of use, creative experience, user impression, and musical practice for evaluation of usability and user experience. Subsequent steps outline the detailed methodologies usable for assessment.}
	\label{fig:flowchart}%
\end{figure}

Therefore, the goal of this article is to provide an  accessible, interdisciplinary overview on current empirical and quantitative approaches to the evaluation of generative systems in music. \alex{Figure~\ref{fig:flowchart} gives a summary and accessible flow chart of the approaches presented.} The article provides an in-depth discussion of evaluation targets and methodologies for the assessment of the output of generative systems as well as the user interaction with such methods (rather than other assessment targets such as sociological implications, etc.). In order to do so, we
\begin{inparaitem}[]
    \item   first introduce a comprehensive overview of the dimensions or targets to be evaluated in Sect.~\ref{sec:targets},
    \item   followed by a description of methodologies and metrics to evaluation of system output and user interaction in Sects.~\ref{sec:output} and \ref{sec:hci}, respectively.
    \item We conclude with a discussion on challenges and future directions in Sect.~\ref{sec:challenges} and final remarks in Sect.~\ref{sec:conclusion}.
\end{inparaitem}

\section{Evaluation targets}\label{sec:targets}
The main goals of evaluating a machine learning system are
\begin{inparaenum}[(i)]
    \item   determining whether the  system works as intended and to what degree and 
    \item   comparing it (quantitatively) with other systems.
\end{inparaenum}
These goals, however, can have many facets as an evaluation can focus on different targets. \citet{wang_controllable_2024} group evaluation targets into ``data-quality evaluation'' and ``property-controllability evaluation,'' the latter focusing on specific output properties that are implicitly or explicitly controlled. \citet{pasquier_introduction_2017} lists the following aspects of a generative system to be assessed: quality, creativity, believability, complexity, robustness, and reliability.
In this work, we propose to group the main evaluation targets into \textit{system output} and \textit{model use} while acknowledging that there are aspects of the model itself and the process of creation that could be subject of evaluation as well.

One of the most common assessment targets is the \textit{system output}. Although the quality of the output is the arguably most intuitive criterion to evaluate, there are several aspects of quality ranging from artistic quality to perceptual audio quality, as well as confounding influences that make the evaluation of output quality a potentially challenging endeavor. 
To give an example of such confounding influences, we can easily imagine an inexperienced listener confusing the artistic quality of a piece of music with the audio quality  of its rendition or the immersiveness of the recording if asked for quality. 
Other dimensions of the output to be evaluated beyond quality include the diversity and originality of the output or how well certain properties of the generated output (e.g., style, rhythmic complexity, or instrumentation) match expectations. 

\alex{Given that} music is a fundamental form of human creativity \cite{kozbelt_evolutionary_2019}, it is also critical to evaluate \emph{model use}~---~how easy to use, enjoyable, and engaging models are for people when they make music\corey{,} whether they be hobbyists, music students, or professional musicians. A model may produce high quality output, but if it is unusable then it has little value for making music. Evaluating the use of models draws on research in Human-Computer Interaction (HCI) \cite{preece_interaction_2015} to assess whether a model is, e.g., usable by people \cite{nielsen_usability_1994}, or \strikethrough{whether it} provokes \strikethrough{feelings of} surprise \cite{caramiaux_explorers_2022} or reflection \cite{ford_towards_2023}.


\subsection{System output evaluation}

    As mentioned above, the most common target for evaluating a generative system is its generated output; the main point of most generative systems is, after all, to produce high quality output or at least output that matches expectations. In this case, the system can be treated as a black box for evaluation as knowledge of internal processes of the system is unnecessary \cite{collins_analysis_2008}.

    Different generative music systems may produce a variety of output formats, which in turn might require different evaluation methodologies and metrics. The output can vary from a single note or sound as generated by synthesizers, a single-voice melody, a polyphonic or multi-voiced musical snippet, to a complete piece of music conforming to structural and other musical expectations. Furthermore, the output may~---in addition to the basic score-based information such as rhythm, harmony, melody---~contain performance information such as tempo and micro-timing, expressive intonation, and dynamics.
    
    The output can either be in an audio format such as PCM \cite{oliver_philosophy_1948}, or in a symbolic format such as MIDI \cite{midi_2023} or MusicXML \cite{good_musicxml_2001}. Note that while any symbolic output might or might not contain (partial) performance information, an audio signal as a ``physical rendition of musical ideas'' \cite{lerch_software-based_2009} automatically contains performance information.    

    The evaluation of the quality of these output signals can be as multi-faceted as different system design goals. This section aims at introducing the main directions of inquiry  for the evaluation of aesthetic quality, the audio fidelity, the originality of the output, and its semantic relevance.
    
        
    \subsubsection{Aesthetic {and musical} qualities}\label{sssec:aesthetics}
        Despite the fact that the assessment of aesthetic quality is possibly the most commonly stated evaluation target in the literature, 
        it arguably is the most challenging to operationally define.
        While aesthetic quality (as applied to a musical artifact) is a catch-all phrase encompassing many attributes such as balance, complexity, novelty, etc. \cite{brattico_subjective_2009}, it is most commonly evaluated only in the singular dimension of subjective preference, i.e., how much a listener ``likes'' a piece of music, and/or how ``interesting'' it is \cite{dervakos_heuristics_2021}.
        However, individual differences\alex{, particularly those that arise from the acquisition of musical expertise,} are known to impact such aesthetic judgments \strikethrough{as well, in particular, those that arise from the acquisition of musical expertise} \cite{pearce_effects_2015}.

        In order to compare across studies, it is important that researchers explicitly operationally define the traits they aim to measure, not only for themselves, but to potential participants as well. 
        For instance, \citet{brattico_subjective_2009} mention the important distinction between \textit{affective} responses (those that induce or modulate emotions or mood), hedonic responses (those that modulate reward; likes and dislikes), and aesthetic responses, which typically refer to inherent style-relevant attributes which lend the artifact beauty, elegance, or coherence. 
        To complicate matters, in addition to style-relevant traits, the medium of creation (e.g., score versus audio), the cultural context (e.g., what is valued inside versus outside the culture), and the caliber or quality of the object or its execution are all important criteria that factor into the equation of evaluating aesthetic goals \cite{galanter_computational_2012}.
        For instance, music generation that outputs a score or transcription may be evaluated on its adherence to various compositional norms, such as adherence to an appropriate vocabulary and grammar, the arrangement and organization of musical ideas, and the use of variation and repetition, to name a few \cite{sturm_taking_2017}.
        On the other hand, the output rendered as an audio recording may be evaluated based on the execution of the performance in relation to parameters such as the authenticity or ``humanness'' of the performance, the expressivity or dynamicism, potentially in addition to factors related to the underlying composition itself. 
        These criteria are \alex{highly} multifaceted \alex{and context-dependent}. 
        For instance, the analysis and assessment of music performance is a  research field in itself \cite{katayose_evaluating_2012, lerch_music_2019, lerch_interdisciplinary_2020}.

        While the assessment of aesthetics typically involves human evaluation, there have been attempts at computational aesthetic evaluation.
        \citet{galanter_computational_2012} gives an overview of such methods through 2012, noting that
       ``computational aesthetic evaluation is an extremely difficult problem,'' and that it frequently ``leads to deep philosophical waters regarding phenomenology and consciousness.''
       Models that have been proposed and used in a musical context, have typically relied on evaluation by adherence to some set of statistical measures or proportions.
      Some AI models that generate an artistic work may be designed so as to implicitly include the goals of such fitness metrics, leading to increased diversity, for example.
       As pointed out by \citet{galanter_computational_2012}\strikethrough{ and others}, however, ``[c]reating evolutionary diversity and dynamics via artificial aesthetics foreign to our human sensibility is one thing. Appealing to human aesthetics is quite another.''
       Other computational avenues for evaluation have included models based complexity~---~frequently drawing on Berlyne's theory of aesthetics, or Shannon information theory \cite{galanter_computational_2012}.

    \subsubsection{Audio quality}
        
         

        The assessment of audio quality, sometimes also referred to as fidelity, is important for a variety of applications, including measuring the quality of audio codecs, the transmission quality of a channel, or the quality of recording or reproduction of audio equipment. Typical factors impacting the audio quality are
        \begin{inparaitem}[]
            \item non-linear processing such as distortion,
            \item changes in the spectral content such as bandwidth reduction,
            \item additive sources such as noise, and
            \item time-varying processes such as gain manipulation or spectro-temporal processing.
        \end{inparaitem}
        When assessing quality in this context, the expectation is that the generated audio signal is free of artifacts and impairments that might negatively impact the human listening experience. 
        The amount (e.g., very little noise vs.\ a lot of noise) of the impairment directly affects the perceived quality of a signal. The estimation of the quality of a signal is easiest in comparison to an identical but unimpaired reference signal. 
                
        A common way to estimate audio quality is through listening studies, with methodologies such as MUSHRA \cite{itu-r_bs1534-3_2015}, which allows to compare the quality of multiple audio signals with respect to a reference signal. 
        Simple objective measures for audio quality such as the Signal-to-Noise-Ratio only have limited perceptual meaning, but there exist objective methods which model perceptual qualities, e.g., PEAQ \cite{itu-r_bs13872006_2006} and ViSQOL \cite{hines_visqol_2015}.

        In addition, attempts have been made to develop reference-agnostic measures of audio quality. This reference-free approach is popular in speech quality assessment. 
        For speech, a clear expectation on quality and intelligibility can be established making the availability of reference signals unnecessary or, at least, less important. In the case of music, however, the artistic and creative use of effects and heavily processed audio question any pre-conceived framework of quality criteria. Thus, the existence of a reference signal is usually deemed necessary for music, as an undesired quality impairment is not always obviously distinguishable from an impairment stemming from artistic intent. The generation of a distorted synthesized sound, for instance, can be perfectly desirable.

    \subsubsection{Originality}
        The originality of the output is a common concern when evaluating a generative system. We interpret originality in three ways. 
        First, originality is a concern with respect to \textit{plagiarism}. Commonly, a generative system is expected to create novel output that does not replicate the training data.
        Second, the \textit{diversity} of the model output should match the diversity of the training data in all relevant dimensions. \alex{This validates the effectiveness of the training.}
        Last but not least, we may want to measure the \textit{creativity} of the system output.
        
        \paragraph{Plagiarism}
            Modern machine learning systems require a potentially massive amount of training data. In most music scenarios, the (partial) reproduction of memorized training data might be regulated by laws. Although systems are generally trained with the intent to generate novel artifacts, they might memorize individual training samples and reproduce them during inference \cite{van_den_burg_memorization_2021, golda_privacy_2024}. When that happens, claims of plagiarism can arise similar to when a human composer copies musical ideas from existing works.
            Examples for  memorized output from generative systems in other fields include 
            \begin{inparaitem}[]
                \item reproduced names and email addresses for  language models \cite{carlini_extracting_2021, huang_are_2022}, and
                \item reproduced images for image generation models \cite{carlini_extracting_2023}.
            \end{inparaitem}
            The analysis of plagiarism in music is more complicated, so cases are not often directly or easily identifiable.
            Therefore, most of the current discussion \strikethrough{of plagiarism} in music focuses on \alex{copyright \cite{lee_ai_2024} and} the use of copyrighted data for training and whether this establishes an infringement of copyright \cite{sunray_sounds_2021, desai_between_2024}. Currently, we see the first lawsuits around this topic reaching the courts \cite{kinsella_time_2024}.
%
            As such potential infringements and associated questions on liability are not necessarily easily addressed and answered, assessing the probability of a generative system reproducing memorized content can be part 
            of an overall evaluation strategy. 
        
        \paragraph{Diversity}
            The generated output of a generative system is expected to show some variety, regardless of use of input prompts or conditioning. Generating only variations of the same piece, for example, is generally not desirable.  
            \citet{banar_quality-diversity-based_2022} point out the importance of measuring diversity of the output in various dimensions as the multitude of musical properties cannot be captured in a single dimension. For common machine learning approaches, the diversity of the generated output should match the diversity of the training set. The exact dimensions to measure diversity on, however, are not necessarily easy to define; they might, for instance, include genre diversity, metric and rhythmic diversity, melodic and harmonic diversity, or diversity of instrumentation and timbre.
            To complicate matters, \citet{liu_impact_2024} show that modern audio synthesis systems often display a trade-off between output diversity and quality of the output.

        \paragraph{Creativity}
            Creativity is a notoriously difficult to define concept. 
            As pointed out by \citet{jordanous_standardised_2012}, there are not even standard definitions of creativity within US or UK law, ``despite the need to detect the presence of creativity for legal reasons.''
            Even more stubborn are the issues that arise in attempting to \emph{assess} creativity, as pointed out by \citet{rohrmeier_creativity_2022}:
                ``It is hard to assess something that one cannot define, and this reflects down to the difficulties in evaluating the success of models of general creativity without resorting to the \corey{`}oracle' of human evaluators.'' 
            While Rohrmeier is referring to the assessment of models of general creativity, the same issues arise in the assessment of creative artifacts or model outputs, regardless of the methods of creation \cite{carnovalini_computational_2020}.
            Despite the fact that creativity is recognized as an ill-defined phenomenon, studies continue to attempt to evaluate it, as it remains a crucial component of artistic creation \cite{kvak_towards_2022}.
            However, in many (if not most) modes of musical creation, one is bound by a set of rules or constraints \cite{chu_empirical_2022}. 
            Again, \citet{rohrmeier_creativity_2022} raises the question of creativity in the context of style imitation where such rules and constraints can, in some cases, be extreme:
                ``What do we mean by \corey{`}creativity',
                and how do we relate novelty, innovation or transformation  with the concept? For instance, are models 'creative' that generate jazz lead sheets, chorales in Bach’s style, Indian tabla, or Balinese Gamelan? Is style replication \corey{`}creative'?'' 
            As pointed out by \citet{agres_evaluation_2016}, creativity \emph{can} be evaluated in a limited sense in a highly constrained context~---such as the harmonization of a melody in a strict chorale style---~as the comparison between human and computational ability to solve a set of problems. This general logic can be extended to theoretically any aspect of creativity. 
            However, even in only evaluating the artifact of a style-imitation task, this procedural or problem-solving definition becomes a slippery slope towards pure determinism, which has been argued to be the opposite of (or, at least, hindering to) creativity \cite{rohrmeier_creativity_2022, zimmermann_creativity_1996}.
            In addition, many attempts to define creativity (e.g., \cite{boden_creativity_1998, jordanous_standardised_2012, cope_computer_2005, ritchie_empirical_2007}) commonly include the notions of value (either aesthetic or utilitarian), the combination or connection of ideas or phenomena, and exploration and transformation within some conceptual space, all of which are missed in the constrained problem-solving definition of style imitation described above.

            Nevertheless, numerous scholars in computational creativity and generative AI agree that, despite the inherent difficulties, assessing creativity~---and having standard, scientific methods for doing so---~is crucial for the field to grow and improve \cite{jordanous_standardised_2012, ritchie_empirical_2007}.

    \subsubsection{Conditioning}
        A common approach to the evaluation of generative systems is to compare specific properties and characteristics of the generated output with the expectation. A system targeting, for instance, the generation of chorales should  not output symphonic music, even if the generated music were aesthetically pleasing and original. 

        Some of the conditioning characteristics are implicitly defined through curating the training data prior to training the system. These could include, e.g., musical genre or style \cite{wang_style-conditioned_2024}, pitch, sonic quality \cite{engel_neural_2017,engel_gansynth_2019}, instrumentation, length, complexity, and mood \cite{godwin_evaluating_2021}.
        Other characteristics can be controlled explicitly either through input conditioning \cite{tan_music_2020} or regularization \cite{pati_attribute-based_2020}. Examples for such characteristics are rhythmic complexity \cite{pati_latent_2019} and arousal \cite{tan_music_2020}.

        Evaluating input conditioning is, in many cases, methodically relatively straight-forward, as the input specifies a target value that the output has to match. If the property can be quantified and measured from the generated output, it can be directly compared with the target value.
        Evaluating the characteristics implicitly specified through the input data can be more challenging as the number and relevancy of properties might not be known, drawing parallels to the evaluation of musical qualities introduced above. Any assessment poses the challenge of identifying a meaningful and complete set of descriptors indicative of the data characteristics to assess. 

        Thus, this direction of evaluation~---focusing on individual output properties and whether they match the user input or training dataset properties---~is mostly useful for the verification that the training process was successful. The evaluation of individual properties allows a very targeted quantitative validation of the semantic relevance of the output with respect to certain, usually narrowly defined characteristics. However, measuring and interpreting specific characteristics of the output as proxies for an overall assessment of the output is, at best, questionable. Any set of characteristics represents only a small subset of a possibly infinite number of musical characteristics, and thus can only give a snapshot of one facet of the generated output. It is particularly problematic if the measured property was explicitly used as a training target or loss function, cf.\ Goodhart's law: ``When a measure becomes a target, it ceases to be a good measure'' \cite{strathern_improving_1997, goodhart_goodharts_2015}.
				
				\alex{A special, but recently very popular form of conditioning is the text prompt. Unlike the evaluation of other conditionings, measuring the (perceptual) alignment of the text prompt with the musical output is challenging due to the undefined structure and terminology or the prompt and the multitude of potentially impacted musical properties. Given these challenges, most evaluation strategies focus on global measures of fit \cite{huang_aligning_2025, grotschla_benchmarking_2025, liu_musiceval_2025}.}

\subsection{Usability \& user experience}

    Outputs from generative music systems are used and appreciated by people from musicians to audiences. In this section, we focus on evaluation of people's direct interaction with generative systems. Viewing this through a HCI \cite{preece_interaction_2015} lens we refer to the people as users who interact with the generative system through a real-time user interface.
    Unlike evaluation of system outputs using listening tests described above, evaluation here is concerned with understanding the interaction between the user and the generative model --- the human-in-the-loop. This aligns with recent Human-Centered AI discourse \cite{shneiderman_human-centered_2022,ozmen_garibay_six_2023} advocating for the use of HCI methods to evaluate and inform the design of AI systems which balance automation with human control. There are two main aspects of interacting with computers, and AI models specifically, that are usually evaluated\corey{:} the \textit{usability} and the \textit{user experience}. 

            The assessment of \textit{usability} asks how easily the generative system is to use \cite{nielsen_usability_1994}. Unlike evaluating the output of the generative system, usability is concerned with how easy the generative system is to control and to understand \cite{amershi_guidelines_2019}. Usability evaluation is often best situated within the wider socio-technical system of use \cite{duh_usability_2006}. Within creative practice, the usability of the system will impact its use and uptake and whether the system is even accepted in a music making context. For example, a generative system may produce aesthetically pleasing outputs, but if it is not usable or controllable it will be less likely to be used in music making practice 
            \cite{louie_novice-ai_2020}. 
        
            The assessment of \textit{user experience} includes collecting subjective and experiential responses to using the generative system \cite{norman_definition_2016}, focusing on evaluating the experience of interacting with the system. This may be, for example, an evaluation of people's affective responses to interaction with the system or an aesthetic evaluation of how the use of the AI relates to music making practice. A person's experience might include hyper-awareness, anxiety, or feelings of control over a situation \cite{csikszentmihalyi_flow_1990}. It might also include feelings of confusion or creative failure when making music \cite{hazzard_failing_2019}, or feelings of surprise when finding unexpected discoveries in AI-generated content \cite{caramiaux_explorers_2022}. In essence, the evaluation of user experience is about assessing people's subjective feelings of using a system.
    
    It is important to note that usability and user experience are interrelated in complex ways --- a system does not necessarily need to be usable to be enjoyable and rewarding to use (e.g., computer games purposefully introduce challenge and frustration to the user experience yet offer a rewarding user experience \cite{blythe_funology_2004}). We see this, for instance, with traditional musical instruments which require years of time and effort to learn, yet this challenge becomes intrinsically rewarding \cite{chaffin_general_2004}.

    

\section{Evaluation of system output}\label{sec:output}
The previous Sect.~\ref{sec:targets} established \textit{what} should be evaluated: the evaluation targets; Sections~\ref{sec:output} and \ref{sec:hci} review \textit{how} these targets can be assessed, discussing methodology and metrics.
For evaluating the generated output of a system, we first discuss subjective evaluation through listening experiments. Then, we describe objective approaches to assess the quality of the output.

\subsection{Subjective evaluation}\label{sec:subjective_eval}
    
    A large problem in the evaluation of system output is that, while one could argue that \textit{style imitation} could be measured fairly objectively without the intervention of human opinion, nevertheless, the evaluation of the utility, aesthetics, creativity, or ``humanness'' is most aptly carried out by a human observer as these are all, essentially, value judgments \cite{carnovalini_computational_2020, schubert_creativity_2021}.
    However, given that the definitions of these traits (utility, aesthetics, etc.) are subjective, largely personal, and subject to contextual information, it remains a significant challenge to design unbiased subjective metrics that would function in an ``all purpose'' manner.
    
    The most typical method for evaluating music generation to date has been by asking listeners \cite{yang_evaluation_2020}. 
    Unfortunately, there are no standardized approaches to the subjective evaluation for almost any Music Information Retrieval (MIR) task or product, including AI-generated music.
    The most common methods for assessing generative music include Turing-style tests~---designed to evaluate whether the AI-generated music can pass as human-generated---~or ``preference tests,'' surveys or experiments assessing aesthetic quality, musicality, and originality. 
    In the ensuing subsections we review these common approaches to subjective evaluation, including both the methodologies themselves as well as the implicit and explicit criteria being evaluated.
    
    \subsubsection{Turing tests}
        Turing-type tests~---where a listener must identify whether a musical selection was made by a human or AI---~remain one of the most common forms of subjective evaluation \cite{hernandez-olivan_subjective_2022}.
        In its most basic form, the Turing test is a useful metric in the sense that the method is simple (i.e., typically binary forced selection), and that it offers a \textit{theoretically} unbiased subjective evaluation by, in principle, implicitly evaluating a model's output according to a scale of ``humanness.''
        If a listener cannot tell apart machine from human-generated musical output, then the implication is that the machine is ``at least as good'' as a human.
        However, this logic only follows under the right conditions, which may not be met in small-scale, ad-hoc experiments.
        For example, what makes a Bach chorale different from another piece of music from the classical genre may be non-evident to a lay listener; similar arguments can be made  for a jazz solo.
        In other words, recognition of the norms of a particular musical style typically takes at least a small degree of expertise \cite{pearce_effects_2015}. 
        Another consideration is the material used in the test itself. 
        Since not \textit{all} of the output material can be evaluated, only a small subset of the model output is used in the test.
        However, depending on how this material is selected, this selection may not \alex{adequately represent} the model output overall. 
        In addition, inferential statistical tests designed to test an alternative hypothesis against a null hypothesis, such as t-tests, are commonly used to evaluate the outcome of a Turing test. 
        Yet typically the desired outcome is that the null hypothesis (i.e., no difference) is actually supported.
        This inappropriate use of such a test will ``bias'' towards supporting a null hypothesis  is compounded by studies that rely on small sample sizes or exhibit high variance.
        %
        The use of Turing tests in evaluating AI output has been criticized for a variety of reasons, most notably for its lack of sophistication, and for the tool being repurposed for something other than what it was intended for \cite{hernandez-orallo_twenty_2020}, which was as an evaluation of intelligence and not of aesthetics.
        It is important to note is that the Turing test conflates \alex{indistinguishability} (between human and computer) with aesthetic or creative success. 
        It is easily conceivable that a model trained on student input, for instance, would ask a listener to disambiguate between two samples both representing `amateur.'
        In this case, a lack of ability to discriminate between the samples does not conclude that the output is of high aesthetic or creative value.
        Finally, an inherent problem with using a Turing test as a subjective evaluation metric is that it inherently over-rewards imitation over creativity \cite{pease_impact_2011}.
        Nevertheless, if used appropriately, the Turing test \textit{can} be used to assess model output, provided that these considerations are taken into account and that the test is primarily used to support a conformity to a baseline standard of a specific musical style, rather than an indication of musically or aesthetically valued output.
        
%

    \subsubsection{Preference tests}
        Other forms of subjective evaluation involving human ratings inevitably fall under the broad category of ``perceptual preference tests,'' where listeners rate their perception of various artifacts according to several criteria, such as aesthetic quality, creativity, musicality, fidelity, etc.
        As mentioned, however, since there are no standardized testing practices for subjective perceptual tests, the questions, the scales used, and the experimental designs all differ widely from study to study (e.g., \citet{chu_empirical_2022}, p.306). 
        The most common criteria evaluated in preference tests include the overall quality, preference or enjoyment, stylistic appropriateness, complexity, coherence, aesthetic response or ``interestingness,'' and musicality; most commonly evaluated on a Likert-type scale \cite{likert_technique_1932}. 
        Despite the wide variety of criteria and survey designs, several scholars have nevertheless made commendable efforts towards breaking down these complex properties in a way that could help impose some common criteria or benchmarks in the design of subjective evaluation practices.
        For example, \citet{chu_empirical_2022} reviewed 40 music generation studies that included a subjective evaluation component and reduced the various criteria into the eight categories
        \begin{inparaitem}[]
            \item \textit{Overall}, 
            \item \textit{Melodiousness}, 
            \item \textit{Naturalness}, 
            \item \textit{Correctness}, 
            \item \textit{Structureness}, 
            \item \textit{Rhythmicity}, 
            \item \textit{Richness}, 
            \item \textit{Creativity}.
        \end{inparaitem}
        Interestingly, the authors imply that \textit{Creativity} was not included as a subjective criterion in any of the studies reviewed, but the authors added this eighth criterion as they felt that ``prior research emphasize[d] the role of AI to boost human creativity in music composition.''
        Unfortunately, however, as presented, these criteria appear difficult to isolate as, for example, the criterion ``melodiousness''~---defined by the question ``are the music notes' relationships natural and harmonious?''---~appears to overlap with the criterion of ``naturalness,'' which asks, ``how realistic is the sequence?'' 
        
        The measurement of audio quality of a signal can be treated as a special case of preference tests with established procedures and methodologies. Audio coding is one of the main fields where the measurement of (perceptual) audio quality plays a crucial role and has driven the standardization of procedures to improve replicability of results. Thus, standards have been introduced that regulate not only the general methodology of listening tests but also number, selection and training of the listeners, selection of audio stimuli, properties of the reproduction equipment, as well as other factors such as room acoustics \cite{itu-r_bs1116-3_2015}. 
        Two standards are most commonly followed for determining musical audio quality: ITU-R BS.1116 for high quality signals and ITU-R BS.1534 for medium quality signals. Both require a reference signal to be present, and rate the quality on a five point scale, although the scales are defined  differently to accommodate for the different targets. In both cases, a high rating is indicative of higher audio quality.  
        BS.1116 \cite{itu-r_bs1116-3_2015} is a so-called double blind, triple stimulus test with hidden reference, where two signals are presented alongside the reference signal and one of the two presented signals is a hidden reference signal. The listener then rates the two signals on the five point scale. 
        BS.1534 \cite{itu-r_bs1534-3_2015}, also referred to as MUlti Stimulus test with Hidden Reference and Anchor (MUSHRA), follows a similar methodology, but adds an ``anchor'' signal that established a comparison point at an easily understandable and reproducible quality level. The methodology allows for multiple quality-impaired stimuli at the same time, and thus generally needs fewer participants to obtain statistically significant results than the more speech-focused Mean Opinion Score (MOS) methodology \cite{itu-t_p800_1996}. 
        MOS methods are popular in speech methodology \cite{itu-t_p800_1996} and have been adopted for the field of generative audio in general \cite{agostinelli_musiclm_2023}. Generally, MOS survey methodology is considered flexible, given that it is not always necessary to provide a reference signal. For instance, the Absolute Category Rating specification of the ITU-T P.800 standard \cite{itu-t_p800_1996} does not require a reference signal (as opposed to the Comparison Category Rating specification). The scale for rating can also vary between a five point scale and a seven point scale.

\subsection{Objective evaluation}

    Aesthetics and the impression of artistic quality are inherently subjective, and thus hard~---or even impossible---~ to approximate objectively. On the other hand, it has been argued that subjective results might not be trustworthy and that we must look beyond ``human opinion for evaluation of computational creativity''  \cite{oneill_limitations_2017}. Whichever side is taken, an evaluation based on computed objective metrics is not meaningless. For instance, \citet{yang_evaluation_2020} show that contemporary generative models can fail to properly model the distribution of even low-level musical properties such as pitch range and count, a result that has been substantiated for an extended set of properties by \citet{banar_systematic_2022}. Thus, metrics for such musical qualities can be helpful in analyzing how well the statistics of the generated output match the statistics of the training data. Furthermore, methodologically sound and properly executed listening studies cannot easily take place after each development iteration to measure progress. Taking into account the other advantages of objective metrics such as perfect reproducibility, objectivity, and scalability to large amounts of data, there is a need for objective measures to assess the output of generative systems.
    There is, however, a risk involved in applying objective metrics to music; \citet{goguen_musical_2004} point out that ``a common approach is to 'bracket' or exile the qualitative aspects, and concentrate attention on aspects that are reducible to scientific analysis.''

    \subsubsection{Methodology}
        
        \begin{figure}
            \centering
            \includegraphics[width=.9\linewidth]{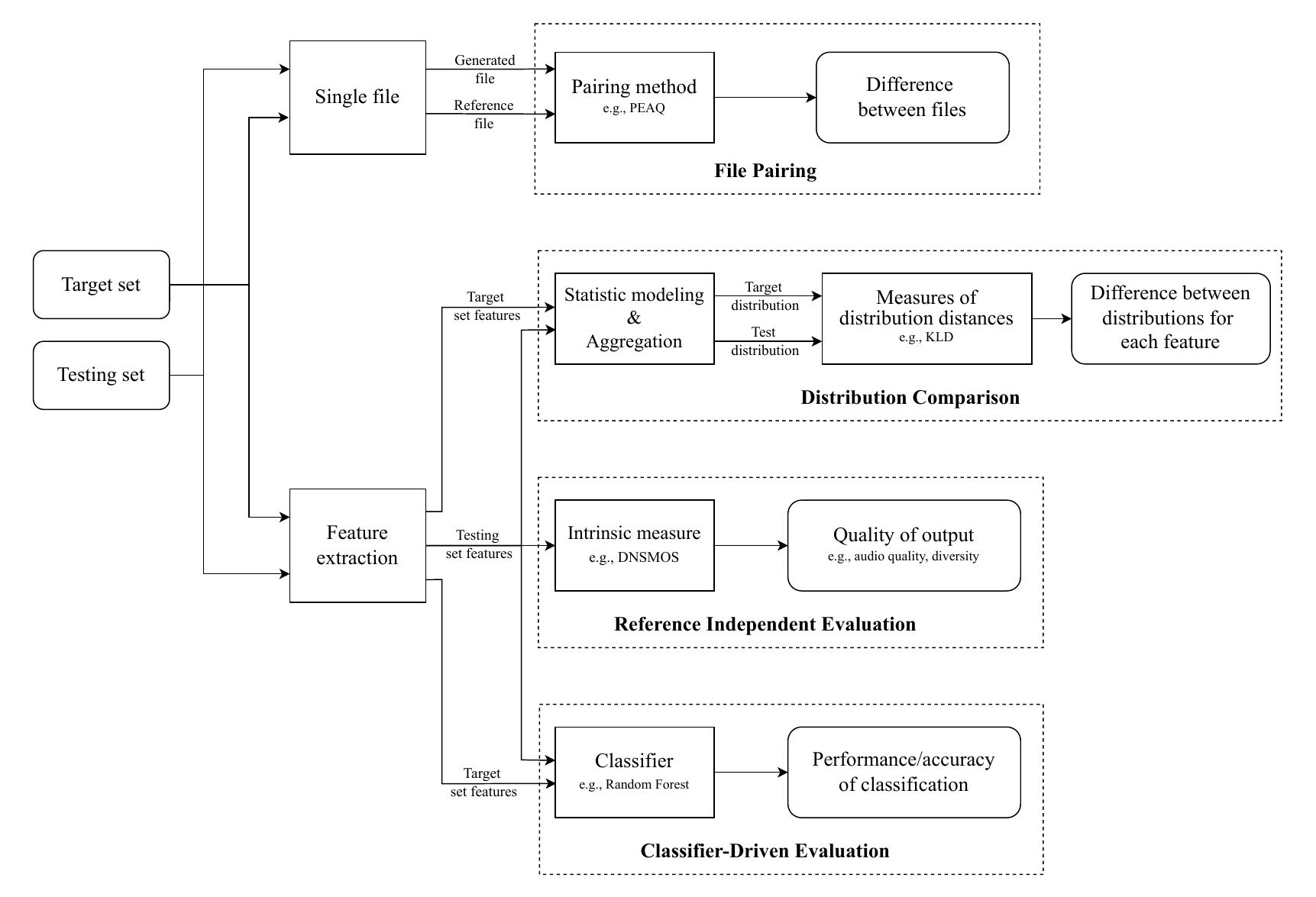}
            \caption{Overview of the evaluation methodologies for generative music systems
            , including both reference-dependent and reference-independent approaches.}
            \Description{The evaluation methodologies structured into a coherent framework that integrates different approaches.}
            \label{fig:objective}
        \end{figure}

        The approaches to the objective evaluation of music generation systems are diverse, as depicted in Fig.~\ref{fig:objective}. Objective evaluation methodology for generative systems typically involves either estimating a score that reflects a specific quality characteristic or using algorithms to compute differences between generated outputs and target benchmarks. In both symbolic and audio generation, the methodologies can be broadly categorized into several distinct types, each serving a different purpose in the evaluation framework. These categories include: 
        \begin{inparaenum}[(i)]
            \item the reconstruction error or match to reference signal,
            \item a comparison of distributions, 
            \item reference-agnostic tests, and 
            \item classifier-driven evaluation.
        \end{inparaenum}

        \paragraph{File pairing \& reconstruction}
            This method compares the generated output directly to a matching reference. Typical use cases are masked music generation (also referred to as music in-painting) \cite{pati_learning_2019} or measuring the reconstruction error of auto-encoder setups. Thus, this approach allows for direct comparison between the generated output and a known reference but only under the limiting assumption that there is \textit{exactly one} correct output. While this assumption allows for a mostly straight-forward assessment approach given a metric (e.g., the Mean Squared Error (MSE) or an edit distances, the Signal-to-Noise Ratio (SNR), or psycho-acoustically motivated metrics such as PEAQ \cite{itu-r_bs13872006_2006}), it is not suited to many typical evaluation scenarios for generative systems where a unique correct output does not exist.

        \paragraph{Comparison of distributions}
            An alternative to comparing files by pairs is to compute distributions of certain characteristics or descriptors across many files to compare a testing (generated) and target (training) set. It is known that generative models are primarily trained to learn characteristics over a training set and generate output replicating those characteristics. Thus, the expectation is for the test distribution to be similar to the target distribution of various descriptors. The target distribution is typically human-composed music. 

            In statistics, a popular distribution distance used to compute differences between distributions is the Kullback-Leibler Divergence (KL-Divergence), which computes the difference between two distributions 
            over the same sample space. The distance measurement is non-symmetric, i.e., the distance from A to B does not generally equal the distance from B to A. 
            In symbolic music evaluation, statistical measures had already been proposed by \citet{collins_analysis_2008}. The KL-divergence is, e.g., used by \citet{yang_evaluation_2020} to evaluate the difference between two distributions in various descriptor dimensions. In addition to KL-divergence, they propose the usage of ``Overlapping Area'' as an indicator of distribution similarity that is~---unlike the KL-Divergence---~both symmetric and bounded. 
            Another commonly used distribution distance is the Wasserstein distance between two multi-variate Gaussians. Recent work by \citet{guo_automatic_2023} proposes calculating a measure of statistical significance between the training and generated distribution.
            All these distance metrics can be calculated from a variety of descriptors, ranging from basic synbolic score features such as pitch histograms to learned embeddings.

            While a high similarity of distributions for various descriptors is desirable, it does not necessarily allow for a conclusive assessment. On the one hand, it depends on how musically meaningful the descriptor itself is. On the other hand, a system could achieve perfect scores by simply replicating the training set, thus leading to favoring conformity over novelty \cite{herremans_functional_2017, briot_deep_2020}.


        \paragraph{Measures without reference}
            This method involves evaluating the quality of generated music without comparing it to a reference or ground truth. Instead, various metrics are employed to assess aspects such as diversity, novelty, coherence, or aesthetic appeal within the generated set itself, thus providing intrinsic (i.e., non-intrusive) evaluations of the generated music, independent of any external reference.
            In many cases, these approaches either measure relative attributes changes of the output without clear anchor point (e.g., increasing novelty beyond a certain point will decrease output quality) or by establishing a framework of clearly defined quality standards that are generally true (e.g., intelligibility for speech signals).
            %
            For example, \citet{chu_song_2016} evaluated repetitiveness within the same corpus of 100 generated songs by segmenting each generated melody into two-bar units and then comparing these segments without the use of an external reference set. Complementing these intrinsic evaluations, \citet{yuan_chatmusician_2024} developed MusicTheoryBench, a set of college-level music theory and composition questions designed to test LLM-based symbolic generation systems' music knowledge and music reasoning.
    
            When evaluating speech quality, no-reference methods are usually preferred. For instance, \citet{manocha_noresqa_2021} find that methods assessing \textit{audio similarity} are not optimal surrogates for assessing speech quality and recommend no-reference methods instead. These methods attempt to estimate audio quality for a signal directly by predicting how humans might rate a signal. 
            
        \paragraph{Classification-driven evaluation}
            A number of evaluation strategies have been proposed to evaluate the quality of generated output through a classifier, i.e., a machine learning system  categorizing input into pre-defined classes. 
            On the one hand, a classification model can be trained to distinguish between human-composed and generated music. 
            The performance of the classifier in accurately distinguishing between real and generated music serves as an (inverse) measure of the quality of the generated compositions.
            On the other hand, classifier performance can be used as a proxy quality measure in specific domains. A genre classifier, for instance, can be used to assess whether the generated music adheres to the desired style  \cite{jin_style-specific_2020, brunner_symbolic_2018} or is similar to the target style \cite{ens_quantifying_2019}. In generative audio tasks, classifiers are directly used to measure whether the generated signals are capable of matching the labels assigned to signals in the training set. In the case of NSynth \cite{engel_neural_2017}, they use a timbral classifier to evaluate whether the generated signals could be trivially classified as being one of the timbre labels in the dataset.
%
            %
            To augment these approaches, \citet{godwin_evaluating_2021} introduced an innovative variation by incorporating generated samples into the training dataset for the classifier and investigated how such augmentation affects the classifier’s performance, shifting the focus from merely comparing the classification of generated outputs with those of the training data to a more dynamic assessment with adaptability.
            

             Classifiers can also be used as the basis for the Inception Score metric \cite{salimans_improved_2016, nistal_comparing_2020}, which is used to measure ``sample diversity,'' i.e., if the model is capable of generating signals that match the label distribution of the testing dataset.
            \citet{banar_quality-diversity-based_2022} review classification-based methods and argue that while these methods might be useful for post-hoc quality evaluation, the embedding spaces utilized by the classifiers are not optimized in terms of the distances between different classes. This lack of optimization can result in suboptimal performance of these models in distinguishing various types of musical content and is of particular concern if the embedding space is used for a distance-based metric. 
            

    \subsubsection{Aesthetic and musical qualities}
         As pointed out above, objectively assessing \textit{aesthetic quality} presents many challenges due to its subjective nature, especially the challenge of how any objective measure could directly reflect aesthetic quality. 
         \citet{juslin_aesthetic_2023} define aesthetic judgment as the assessment process through which the value of a piece of music as ``art'' is determined based on subjective criteria such as novelty, expressivity, and beauty, which relate to both the form and content of the artwork. They state that aesthetic judgments result from psycho-physical interactions between the music's objective properties and the subjective impressions of the evaluator. Therefore, there are no absolute or universal criteria for aesthetic value, as aesthetic norms are subject to change over time within society. \citet{kalonaris_computational_2018} substantiated this perspective by noting that many experiments and theories on musical aesthetics~---such as concepts of beauty, pleasantness, and well-formedness---~are heavily reliant on context-specific and often arbitrary assumptions about the nature of music, and are thus hard to generalize. 
         Despite this known variability,  attempts towards developing objective evaluation metrics of aesthetic quality have been made, drawing on both traditional music theory and empirical aesthetics principles.

        \citet{kalonaris_computational_2018} categorize the measures of computational aesthetics in music into several distinct types, each grounded in different theoretical frameworks. These include 
        \begin{inparaenum}[(i)]
            \item information and complexity-based aesthetic measures, such as entropy, which quantify the order and predictability within musical compositions,
            \item geometric measures, which assess aesthetic value by analyzing the statistical distribution of musical elements,
            \item psychological measures, such as Gestalt principles of grouping, to understand how listeners perceive and interpret musical structures as a whole rather than merely sums of parts, and 
            \item biologically inspired measures, which explore the social, genetic, and evolutionary algorithms.
        \end{inparaenum}

       Notably, evidence has been presented that many musical parameters (e.g., pitch, dynamics) follow a Zipfian (power law) distribution and that adherence in musical composition to such distributions is more aesthetically pleasing \cite{galanter_computational_2012}. Zipf's law \cite{zipf_human_1965} suggests that the frequency of a symbol within a piece of music should be inversely related to its rank in terms of frequency of use. 
       Computational aesthetic measures relying on predictions from Zipfian distributions have been successfully used to predict human aesthetic ratings such as ``pleasantness'' for both music and visual art \cite{manaris_evolutionary_2003, manaris_developing_2005, galanter_computational_2012}. It should be noted that the concepts of musical ``pleasantness'' and overall aesthetics or ``appreciation'' \cite{pearce_effects_2015} might not be necessarily synonymous, as the latter implies a broader range of emotional and intellectual responses. 
       Another automated computational approach reviewed by \citet{galanter_computational_2012} include using various simulation models based on chromosome behavior that compute a single weighted metric, an ``evolutionary fitness score,'' that rewards some desired behavior or trait.
        
        Berlyne's theory of ``New Empirical Aesthetics'' \cite{berlyne_aesthetics_1971} provides another theoretical framework for understanding aesthetic quality. According to this theory, the appreciation of an artwork correlates with its complexity and its ability to stimulate arousal, with the audience's liking having an inverted-U relationship with ``arousal potential.'' In a practical setting, the measurement of subjective aesthetic quality often includes beauty, groove, originality, complexity, expression, emotion, sound quality, prototypicality, message, and skill \cite{juslin_everyday_2013}.
        
        \begin{footnotesize}
            \begin{table}
\caption{Previously used low-level descriptors used to describe musical qualities.}\label{tab:descriptors}            
\centering
\begin{tabular}{p{.08\linewidth}|p{.08\linewidth}|p{.25\linewidth}|p{.5\linewidth}}
\hline                                                                                \textbf{Type}   & \textbf{Unit}    & \textbf{Name}                              & \textbf{Description}                                               \\
\hline                                                                                                   
\multirow{12}{=}{{Pitch \cite{ji_survey_2023, yang_evaluation_2020, dong_musegan_2018, mogren_c-rnn-gan_2016}}}
				& \multirow{6}{*}{Note}
								   & Total Used Pitches, Pitch Class Histograms  &  number of distinct pitches or pitch classes (often accompanied by entropy)                                         \\
                &                  & In-Key Note Frequency, Scale consistency &  fraction of notes adhering to a scale                                                                                                          \\
                &                  & Pitch Range, Tone Span                    & span between the highest and lowest pitches (usually in semitones)                                                                              \\
                &                  & Consecutive Pitch Repetition               & frequency of repetitions of a pitch                                                                                 \\
                &                  & Pitch Interval Average, N-gram            &  interval between consecutive notes (usually in semitones)  \\
                &                  & Pitch Class Transition                     &  frequency and type of transition between pitch classes                                                                                  \\
                \cline{2-4}
                & \multirow{6}{*}{Sequence}
								& Empty Bars                                 & number or ratio of bars w/o musical onsets                                                                                              \\
                &                  & Sequence Repetition                        & number of repetitions of short sequences                                                                                                                   \\
                &                  & Rote Memorization Frequency                &  frequency of which the generated sample reproduces exact sequences from the training corpus                                              \\
                &                  & Frequency of Pitch Change                  & how often pitches change                                                                                                                     \\
                &                  & Pitch Variations                           &  number of distinct pitches                                                                                                              \\
                &                  & Voice Motion                               & how voices or melodic lines move relative to each other                                     \\
\hline
\multirow{8}{=}{Rhythm \cite{ji_survey_2023, yang_evaluation_2020, mogren_c-rnn-gan_2016, wu_jazz_2020}}
                & \multirow{5}{*}{Note}
                                   & Total Used Notes                           &  total number of notes                                                                                                      \\
                &                  & Note Length Histogram                      &  distribution of note lengths                                                                                                     \\
                &                  & Qualified Note Length Frequency            &  frequency of note durations                                                                                                   \\
                &                  & Average Inter-onset Interval               &  average time between the onset of consecutive notes                                                                                                            \\
                &                  & Note Length Transition                     &  frequency and type of transition between consecutive note lengths                                                                                            \\
                \cline{2-4}
                & \multirow{3}{*}{Sequence}
								 & Rhythmic Similarity between Measures       &  rhythmic similarity between different measures                                                    \\
                &                  & Rhythm Variations                          &  number of distinct note durations                                                                                                      \\
                &                  & Off-beat Recovery Frequency                & how frequently the model can recover back onto the beat after being forced to be off beat                                                                 \\
\hline
\multirow{9}{=}{Harmony \cite{ji_survey_2023, wu_jazz_2020, ens_quantifying_2019}}
                & \multirow{5}{*}{Chord}            
                                    & Chord Duration                             &  length of time a chord is held                                                                                                                                \\
                &                  & Chord Content                              & pitch composition of a chord                                                                                                              \\
                &                  & Chord Vocabulary                           & e.g., number of used chords, chord coverage, number of repeated chords                                                          \\
                &                  & Tonal Distance Average, Histogram         &  average tonal distance between  pairs of adjacent chords                                   \\
                &                  & Chord Tone to Non-Chord Tone Ratio         &  in-chord vs.\ non-chord tones                      \\
                \cline{2-4}
                & \multirow{4}{*}{Overall}
                                    & Melody-Chord Tonal Distance          &  average tonal distance between each melody note and its corresponding chord                                                                                    \\
                &                  & Progression Irregularity                   &  degree of difference in chord progressions between a sample and  templates                                                                         \\
                &                  & Polyphonicity                              &  frequency of simultaneously played pitches                                                                                                  \\
                &                  & Dissonance                                 &  dissonance level of onsets based on their periodicity     \\
\bottomrule
\end{tabular}
\end{table}
        \end{footnotesize}
        Metrics targeting \textit{musical qualities} are commonly distribution-based, meaning that the similarity between a generated and a target distribution is measured for a specific musical descriptor or property. Typically, these descriptors are low-level representations of music \cite{ji_survey_2023} and s%
        ome of these features and distributions might be more musically relevant than others. For example, pitch distributions are commonly applied to MIDI numbers or pitch values (e.g., A4) as opposed to much more meaningful pitch or chord distributions that take key information into account, such as scale degree (the position of a pitch class in a tonal context) or Roman numeral (the function of a chord given its context). Still, basic features such as pitch difference and the ratio of in-key pitches or entropy of pitch and chords \cite{wang_intelligence_2023} continue to be employed in recent works and are sometimes referred to as ``music theory evaluation'' \cite{guo_automatic_2023}. 
        Over time, many descriptors for musical qualities have been introduced \cite{meredith_computational_2016}; \citet{yang_evaluation_2020} proposed a set of simple metrics separately targeting tonal and rhythmic content. Similarly, \citet{garcia-valencia_framework_2021} utilized the concepts proposed by \citet{tymoczko_geometry_2011} to assess qualities like the smoothness of melody transitions, pitch diversity, and the occurrence of local notes to provide an analysis of melodic structure with rhythmic features. \citet{dervakos_heuristics_2021} introduced a framework based on the basic consonance aspects of melodies that allows the construction of a variety of metrics. They used this framework to construct four heuristic properties, polyphonicity, used-pitch-classes per bar, total number of pitches, and the total number of pitch classes, to address fundamental musical properties that are reliable and interpretable. These examples show the variety of descriptors that have been used to model musical properties. Table~\ref{tab:descriptors} provides an overview of common descriptors proposed in the literature.
            
        While there have been many calls for more musically meaningful and relevant objective metrics, many of these descriptors and resulting metrics are highly specialized to certain styles, genres, or musical content. For example, Wu and Yang \cite{wu_jazz_2020} employed specific metrics such as grooving and chord progression to evaluate generated jazz performances. They observed that erratic usage of pitch classes, inconsistent grooving pattern and chord progression, and the absence of repetitive structures contribute to the quality gap between the generated and human-composed jazz samples. Although these metrics appear to be pertinent to jazz, they might not fully capture or evaluate the qualities of music produced in other genres. This specialization of metrics may restrict their broader application across different musical styles and genres and implies that other musical styles might require the design of style-specific descriptors as well.
        
        Furthermore, scholars have pointed out the importance of higher level features such as form and repetition in contributing to the improvement of generative music systems (e.g., \cite{dai_what_2022}). However, specific metrics to measure these higher-level features have not yet been proposed. \alex{Instead, metrics such as the Fréchet Distance (see Sect.~\ref{sec:objective-metrics}) utilizing trained embeddings with opaque or unknown musical meaning (VGGish \cite{hershey_cnn_2017}, CLAP \cite{wu_large-scale_2023}, etc.) have been increasingly applied to music evaluation \cite{gui_adapting_2024,retkowski_frechet_2025,huang_aligning_2025,chung_kad_2025}.}

    \subsubsection{Audio quality}\label{sec:objective-metrics}

        For any generative audio system, assessing the audio quality of the outputs is integral to understand whether the generated signals \strikethrough{would} can be considered high-quality by human listeners. Simple error measurements such as the MSE, the Mean Absolute Error (MAE), or the SNR have been long shown to be ineffective as ``quality'' measurements \cite{karjalainen_new_1985}, so that listening studies are often considered to be the ultimate way to assess audio quality. Regardless, objective metrics for measuring audio quality have been proposed. 
        
        In many cases, where generative systems are trained to synthesize and match the inputs as closely as possible, differences in the presence of artifacts, distortion, noise and bandwidth are measured. This usually assumes that the cleanest reproduction of the signal is free of artifacts, distortion and noise while maintaining an identical frequency bandwidth to the original signal. If the reference signal is available, it is feasible to use metrics that rely on file pairing. 
        Popular reference-based objective methods are 
        \begin{inparaenum}
            \item \textit{Perceptual Evaluation of Audio Quality} (PEAQ) \cite{itu-r_bs13872006_2006}, a method that uses a psycho-acoustic model to compute features representing the difference between a reference and a test signal,
            \item \textit{Signal-to-Noise Ratio} (SNR), \textit{Signal-to-Distortion} Ratio (SDR) and \textit{Signal-to-Artifact} Ratio (SAR), which compute the difference in noise levels, distortion levels, and artifact levels between paired signals, respectively, and, 
            \item Deep Perceptual Audio Metric (DPAM) \cite{manocha_audio_2022}, a  metric that is trained to estimate perceptual quality similarity between paired signals. 
        \end{inparaenum}

        Many of the contemporary approaches to the problem use the aforementioned distribution comparison metrics. By and large, these methods extract neural representations or embeddings for both the target and testing sets. The primary argument for the usage of distribution comparison methods hinges on the notion that minimal divergence between the two distributions is indicative of the two sets being roughly identical in terms of quality. The most popular distribution divergence metric used in evaluating generative quality is the Fréchet Audio Distance (FAD) \cite{kilgour_frechet_2019}, which computes the difference between the two embedding distributions using a Wasserstein distance on VGGish embeddings \cite{hershey_cnn_2017}\alex{, and more recently on a variety of other embeddings, such as PANNs \cite{kong_panns_2020}, CLAP \cite{wu_large-scale_2023}, and Encodec \cite{defossez_high_2023} (cf.\ \cite{gui_adapting_2024,grotschla_benchmarking_2025})}. 
        The perceptual relevance of FAD has, however, been questioned for both audio quality evaluation \cite{vinay_evaluating_2022} \alex{and for music evaluation \cite{huang_aligning_2025}.} \citet{gui_adapting_2024} propose a variation of the FAD for audio-based music quality measurement called FAD$^\infty$. FAD$^\infty$ is an extension of FAD that mitigates the sample-size bias by estimating the behavior of the metric as if it had an infinite number of samples by using Quasi Monte-Carlo integrals. \citet{gui_adapting_2024} \strikethrough{also investigate replacing the VGGish embedding with different embeddings and} report that~---given the right embeddings and mitigating sample size bias---~FAD$^\infty$ can be a useful indicator of quality, a result that is intended to assuage concerns raised by previous results \cite{vinay_evaluating_2022}.
        An alternative to FAD is to use Maximum-Mean Discrepancy (MMD) \cite{binkowski_demystifying_2018, nistal_comparing_2020}, which is a two-sample test computed over the kernel embedding of the training and test data. For more details on the mechanics of MMD, we refer the reader to \citet{JMLR:v13:gretton12a}. Recent work in the image domain has shown that MMD with CLIP embeddings is a better approximator of generative image quality than Fréchet distances \cite{jayasumana_rethinking_2023}. \alex{\citet{chung_kad_2025} propose to adapt the MMD to the Kernel Audio Distance (KAD) to replace FAD.}

        Distribution comparison methods 
        measure if the generator is capable of modeling the underlying distribution that it was trained to generate. However, 
        the perceptual relevance of the result largely depends on the selected feature / embedding space \cite{ananthabhotla_towards_2019, vinay_evaluating_2022}. Unbounded, distance-based metrics also lack interpretability in the sense that even a statistically significant difference in a metric does not necessarily imply a perceptually significant difference.        

        \begin{footnotesize}
            \ifthenelse{\equal{\bNewTable}{false}}
{\newcommand{\perceptual}[0]{\colorbox{blue!30}{Perceptual}}
\newcommand{\isml}[0]{\colorbox{red!20}{Uses ML}}
\newcommand{\distribution}[0]{\colorbox{yellow!20}{Distribution Driven}}

\begin{table*}[ht]
    \tiny
    \centering    
    \begin{tabular}{|l|l|c|c|l|}
    \hline
        Metric & Type & Properties & Domain & Range \\ \hline
        
        PEAQ & 
        \begin{tabular}{@{}c@{}}
 		Paired Reference \& Test\\
 		\end{tabular}
 		& 
 		\begin{tabular}{@{}c@{}}
            \perceptual\\
            \isml
		\end{tabular} & 
		\begin{tabular}{@{}c@{}}
			Agnostic\\
			Codecs
		\end{tabular} &
		-4 to 0 $\uparrow$
		\\
		\hline
		
        NDB/$k$  & 
        Reference and test distribution & 
	\begin{tabular}{@{}c@{}}
            \distribution\\
            \colorbox{red!20}{Uses ML}
        \end{tabular} & 
        \begin{tabular}{@{}c@{}}
        Deep Learning\\Images
        \end{tabular}& 
        0 to $\infty$ $\downarrow$
        \\
        \hline
        
        Mean Squared Error & 
        Paired Reference \& Test & 
        \begin{tabular}{@{}c@{}}
        \colorbox{green!20}{Differentiable}\\
        \colorbox{black}{\textcolor{white}{Distance}}
        \end{tabular}& 
        Agnostic& 
        0 to $\infty$ $\downarrow$
        \\ \hline
        
        Mean Absolute Error & 
        Paired Reference \& Test & 
        \begin{tabular}{@{}c@{}}
        \colorbox{green!20}{Differentiable}\\
        \colorbox{black}{\textcolor{white}{Distance}}
        \end{tabular}& 
        Agnostic &
        0 to $\infty$ $\downarrow$
        \\ 
        \hline
        
        DPAM & 
        Paired Reference \&  Test & 
        \begin{tabular}{@{}c@{}}
        \colorbox{green!20}{Differentiable}\\
        \colorbox{blue!30}{Perceptual}\\
        \colorbox{red!20}{Uses ML}
        \\
        \end{tabular}
        & 
        \begin{tabular}{@{}c@{}}
        Deep Learning\\Speech\\ 
        \end{tabular}& 
        0 to $\infty$ $\downarrow$\\
        \hline

        
        Fréchet Audio Distance & 
        Reference and test distribution & 
        \begin{tabular}{@{}c@{}}
        \colorbox{green!20}{Differentiable}\\
        \colorbox{black}{\textcolor{white}{Distance}}\\
        \colorbox{yellow!20}{Distribution-Driven}\\
        \colorbox{red!20}{Uses ML}
        \end{tabular}
        &
        \begin{tabular}{@{}c@{}} 
        Deep Learning\\Music
        \end{tabular}&  
        0 to $\infty$ $\downarrow$
        \\ \hline
        
        SI-SDR & 
        Paired Reference \& Test & 
        \colorbox{blue!30}{Perceptual} & 
        Agnostic& 
        0 to $\infty$ $\downarrow$
        \\ 
        \hline
        
        SNR & 
        Paired Reference \& Test & 
        \colorbox{blue!30}{Perceptual} & 
        Agnostic &
        0 to $\infty$ $\downarrow$
        \\ 
        
        \hline
        
        
        Inception Scores & 
        \begin{tabular}{@{}l@{}}
        No Reference,\\
        Reference Class/Test Distribution
        \end{tabular}
        & 
        \begin{tabular}{@{}c@{}}
        	\colorbox{green!20}{Differentiable}\\
        	\colorbox{red!20}{Uses ML}
        \end{tabular}
        & Deep Learning & 
        0 to $\infty$ $\uparrow$ \\ 
        \hline
        
        Kernel Distance  & 
        Reference and test distribution & 
        \begin{tabular}{@{}c@{}}
        \colorbox{green!20}{Differentiable}\\
        \colorbox{black}{\textcolor{white}{Distance}}\\
        \colorbox{yellow!20}{Distribution-Driven}\\
        \colorbox{red!20}{Uses ML}
        \end{tabular}
        &
        \begin{tabular}{@{}c@{}}
        Agnostic\\
        Deep Learning
        \end{tabular}& 
        0 to $\infty$ $\downarrow$
        \\ \hline
        
        KL-Divergence & Reference and test distribution &
        \begin{tabular}{@{}c@{}} 
        \colorbox{green!20}{Differentiable}\\
        \colorbox{black}{\textcolor{white}{Distance}}\\
        \colorbox{yellow!20}{Distribution-Driven}
        \end{tabular}
        & 
        \begin{tabular}{@{}c@{}}
        Agnostic\\Deep Learning
        \end{tabular}& 
        0 to $\infty$ $\downarrow$
        \\ \hline
        
        DNSMOS & 

        No Reference &
        \begin{tabular}{@{}l@{}}
        \colorbox{blue!30}{Perceptual}\\
        \colorbox{red!20}{Uses ML}
        \end{tabular} & 
        \begin{tabular}{@{}l@{}}
        Deep Learning\\Speech
        \end{tabular}&
        0 to 5 $\uparrow$ 
        \\
        \hline               
    \end{tabular}
    \caption{A table containing popular metrics and distances that are used to evaluate audio quality. In the range column, $\downarrow$ indicates that lower is better and vice versa} 
	\label{tab:metrics}
\end{table*}
}

{
\begin{table}
\centering
\caption{Commonly used metrics and distances for evaluation of audio quality. In the range column, $\downarrow$ indicates that lower is better and vice versa.}
\label{tab:metrics}
\begin{tabular}{l|l|c|l} 
\hline
\textbf{Type}                                                                             & \textbf{Metric}        & \textbf{Domain}                                                 & \textbf{Range}              \\ 
\hline
\multirow{6}{*}{Paired reference \& test}                                                   & PEAQ                   & \begin{tabular}[c]{@{}c@{}}psycho-acoustic\\model\end{tabular}  & -4 to 0 $\uparrow$          \\ 
\cline{2-4}
& Mean Squared Error     & waveform or spectral                                            & 0 to $\infty$ $\downarrow$  \\ 
\cline{2-4}
& Mean Absolute Error    & waveform or spectral                                            & 0 to $\infty$ $\downarrow$  \\ 
\cline{2-4}
                                                                                          & DPAM                   & trained from listening study                                    & 0 to $\infty$ $\downarrow$  \\ 
\cline{2-4}
                                                                                          & SI-SDR                 & waveform                                                        & 0 to $\infty$ $\downarrow$  \\ 
\cline{2-4}
                                                                                          & SNR                    & waveform                                                        & 0 to $\infty$ $\downarrow$  \\
\cline{2-4}
& ViSQOL & trained from listening study & 1 to 5 $\downarrow$ \\
\hline
\multirow{4}{*}{Ref.\ \& test distribution}                                          & NDB/$k$                & \begin{tabular}[c]{@{}c@{}}Deep Learning\\Images\end{tabular}   & 0 to $\infty$ $\downarrow$  \\ 
\cline{2-4}
                                                                                          & Fréchet Audio Distance & VGGish, \alex{CLAP, etc.}                                                          & 0 to $\infty$ $\downarrow$  \\ 
\cline{2-4}
                                                                                          & Kernel Distance        & trained from labels                                             & 0 to $\infty$ $\downarrow$  \\ 
\cline{2-4}
 & Inception Scores       & trained from labels                                             & 0 to $\infty$ $\uparrow$    \\ 
\hline

\hline
No reference                                                                              & DNSMOS                 & trained from listening study                                       & 0 to 5 $\uparrow$           \\
\bottomrule
\end{tabular}
\end{table}
}
        \end{footnotesize}

        Other notable approaches to estimating audio quality include DNSMOS \cite{reddy_dnsmos_2021} NDB/$k$ \cite{richardson_gans_2018} and ViSQOL \cite{hines_visqol_2015},  which are predominantly speech quality focused approaches. ViSQOL and DNSMOS are speech quality estimation methods trained using results from large-scale MOS listening studies. ViSQOL is a paired metric that uses spectrogram patches and  DNSMOS is a no-reference metric that uses a model trained to predict what a human might rate a signal. NDB/$k$ is a metric originally proposed for images that has been used to evaluate some generative systems such as Diffwave \cite{kong_diffwave_2020}. It clusters the features of the training set using k-means clustering and Voronoi cell partitioning. To compute the metric, the number of statistically different bins or cells are computed.
        Recent work has shown that in speech, similarity is not a reliable proxy for quality and no-reference metrics, such as the ones mentioned above, might be better at estimating quality \cite{manocha_audio_2022}.

        
        Table~\ref{tab:metrics} shows an overview of commonly used objective metrics for audio quality assessment.
        

    \subsubsection{Originality}\label{ssec:originality}
        Originality is often considered a key characteristic of a generative system, however, only few quantitative measures have been proposed despite calls for work on, e.g., quantification of output diversity \cite{epstein_art_2023}. As mentioned above, we understand originality to be comprised of \textit{diversity}, \textit{novelty / plagiarism}, and \textit{creativity}.

            While \textit{diversity} is implicitly evaluated by commonly used metrics for image generation such as the Inception Score and the Fr\'echet Inception Distance, they do not differentiate the diversity aspect from the quality/fidelity aspect. Other quantifiable metrics on diversity have been pre-dominantly proposed for GAN-based generative systems \cite{borji_pros_2022}. However, as \citet{gulrajani_towards_2019} point out, many of the existing metrics can be tricked by the model simply memorizing the training data. Thus, they propose so-called neural network divergences (NNDs), measured through the loss of a neural network trained to distinguish between generated and training data, and show that NNDs can serve as meaningful measures of diversity.
            \citet{alaa_how_2022} propose to measure diversity using a metric they refer to as $\beta$-recall, measuring the fraction of real samples covered by the most typical generated samples.
            In the context of music, \citet{yin_good_2021} propose the ``Originality Score'' to  quantify how original a set of musical pieces is and compare the originality scores between training and generated data to understand if the originality of the generated data matches  expectations.

            \citet{sturm_artificial_2019} note that for music~---unlike for text---~automated \textit{plagiarism} detection does not reliably work, and worse, no clear standards exist about what type and amount of alterations make a piece of music novel as opposed to plagiarized. Thus, approaches such as the Authenticity metric proposed by \citet{alaa_how_2022}, estimating the probability of a generated sample being copied from the training data, are only of limited use.
            Still, approaches to objectively measure plagiarism have been proposed, usually framed as a music similarity task. Most of the seminal work can be grouped into
            \begin{inparaenum}[(i)]
                \item melodic similarity measures based on symbolic input and
                \item general audio-based similarity measures.
            \end{inparaenum}
            To measure \textit{symbolic melodic similarity}, a variety of metrics have been proposed. Most of them are based on some form of sequence similarity \cite{park_music_2005, mullensiefen_court_2009, wolf_perception_2011, cason_singing_2012, de_prisco_music_2017, de_prisco_computational_2017} or vector-based similarity measures hand-designed \cite{malandrino_adaptive_2022} or trained as end-to-end systems \cite{park_music_2022, lu_melodysim_2025}. Some of these algorithms have been tested against real-world court decisions, however, the test set sizes are necessarily small and external validity is hard to verify.
            \textit{Audio-based similarity} \cite{lerch_introduction_2023} is a multi-dimensional problem at the risk of confounding dimensions of score similarity (melody, harmony, rhythm, etc.) with performance similarity (tempo, playing techniques, etc., but particularly timbre). Systems for audio-based plagiarism detection have been proposed to use Non-Negative Matrix Factorization to decompose the audio \cite{de_plagiarism_2012}, MFCC vector-based representations of audio \cite{suneja_comparison_2015}, similarity measures inspired by audio fingerprinting approaches \cite{borkar_music_2021, lopez-garcia_proposal_2022}, \alex{or utilizing pre-trained embeddings for similarity measurement \cite{batlle-roca_towards_2024}}.
            Commonly, however, plagiarism is understood in terms of score similarity (with the very prominent exception of sampling, where audio is copied and mixed into a new musical artifact \cite{gururani_automatic_2017}). A study into the contributing factors of court cases on plagiarism has been presented by \citet{yuan_perceptual_2023}.
        
            The evaluation of \textit{creativity} poses an unsolved problem. \citet{jordanous_standardised_2012} notes the longstanding lack of attention to evaluation and lack of evaluation standards in the computational creativity community due to ``difficulties in defining what it means for a computer to be creative; indeed, there is no consensus on this for human creativity, let alone its computational equivalent.'' 
            Indeed, in a survey of 75 papers on computational creativity, the author found that in one third of the papers, `creativity' was not mentioned, and that only one third of the papers actually attempted to evaluate the creativity of their systems. 
            Moreover, ``[o]ccurrences of creativity evaluation by people outside the system implementation team were rare.''
            Judging from the scarcity of proposed evaluation standards since the publication of this paper~---despite the fact that it has been over a decade since its publication and that creative AI is presently in its heyday---~it seems that this issue continues to persist.
            As mentioned above, one can distinguish between the evaluation of a computational system itself, and such a system's output (i.e., ``the product/process debate'' \cite{jordanous_standardised_2012}). 
            However, since the output is restricted or constrained by the system, it can be helpful to conceptualize a combined methodology for the evaluation of both the system (in terms of its creative potential) and its output.
            \citet{jordanous_standardised_2012} presents such a framework, referred to as ``SPECS'' (Standardised Procedure for Evaluating Creative Systems).
            The SPECS method~---which is not only well recommended \cite{oneill_limitations_2017}, but unlike other proposed models has actually been used in practice \cite{ostermann_evaluating_2021,plucker_assessment_2010}---~asks the creator/evaluator to adhere to a specific working definition of creativity that includes two or more ``components'' that are evaluated independently according to some standard, such as skill, novelty, value, etc., and compared against an appropriate system/output. 
            These evaluations may be quantitative or qualitative, however, in this case, both still involve significant investment of time and resources into evaluation with human subjects.
            While other models and frameworks have been proposed for attempting to theoretically quantitatively assess creative output, such at the FACE and IDEA models \cite{pease_impact_2011}, to our knowledge, \strikethrough{there remain} no adopted standardized practices or procedures \alex{currently exist} for assessing creativity without human intervention\strikethrough{ at this time}.

\section{Evaluation of usability \& user experience}\label{sec:hci}

In this section we describe Human-Computer Interaction (HCI) methods and techniques that have been used to evaluate user interaction with generative music models.  
These methods and techniques share aims with Human-Centered AI \cite{shneiderman_human-centered_2022,ozmen_garibay_six_2023} to research, design, and evaluate AI systems from a human-centered or user-first perspective. 
Below, we first introduce HCI methodologies for evaluating AI music systems. We then describe data collection methods for understanding the user experience of generative AI including quantitative and qualitative approaches. \nbk{Table \ref{tab:hci_evaluation} closes this section by summarizing key HCI evaluation approaches discussed.} 
It is important to note that there are no de-facto standards defining which data collection methods are used for which evaluation methodologies. Instead, selection of data collection and methodology is based on best practice in the field and individual HCI practitioner's skills and experience. 


\subsection{Methodology} 

HCI research has traditionally focused on functional aspects of user interaction such as the usability of a system \cite{nielsen_usability_1994}, whilst later waves of HCI placed more focus on subjective qualities of users' experience when interacting with computers \cite{bodker_third-wave_2015}. 
Approaches to HCI evaluation of generative music systems draw on both functional (usability) and experiential (user experience) forms of HCI evaluation. Broadly, evaluation methods and can be split into controlled experiments which are more objective and typically suited to exploring the usability of a system, and more subjective and ecologically valid (meaning applicable to real-world practice) approaches which are more suited to studying experiential aspects of generative music systems \cite{bryan-kinns_guide_2023}.

\subsubsection{Controlled experiments}
Controlled experiments place participants in distraction free environments such as a research lab, where they are set a number of musical tasks to complete with an AI system in a constrained amount of time. 
In this setting, typically two or more versions of an AI system are used to allow comparison between versions, often referred to as A-B testing. The tasks undertaken might be open-ended (e.g.,  write a piece of music) or more specific (e.g.,  harmonize a given melody) depending on the model and the evaluation goals. For example, \citet{suh_ai_2021} and \citet{louie_novice-ai_2020} asked participants to compose music for a fictional character from a game, while \citet{frid_music_2020} asked participants to create music for a video based on an example song.
In \citet{louie_novice-ai_2020}, two versions of a music making interface for harmonizing melodies were tested~---one with and one without AI-steering tools---~to allow for comparison of the effect of the AI. In this case the tool without AI features can be referred to as a baseline. Alternatively, different forms of interaction and participation might be tested with the same system, e.g., \citet{addessi_flow_2015} compared their AI music improvisation system when used by individual children vs.\ groups of children.

\subsubsection{Online evaluation}
Online evaluation settings can be helpful for evaluating generative AI systems across a large sample size. For example, \citet{ben-tal_how_2021} \strikethrough{examined} \corey{evaluated} a version of the \emph{FolkRNN} generative AI model hosted online, where they were able to examine how people generated content with FolkRNN serendipitously and how they tweaked values to modify and curate outputs. Typically, generative AI systems are either hosted online for user interaction or available as a download to users. Audiences across the globe can then be reached using survey platforms such as Prolific,\footnote{\url{https://www.prolific.com/}, last accessed: Jun 25, 2024} and filtered for characteristics such as nationality or technical skills. Whilst online settings allow for large numbers of participants, they lack the rigor of controlled experiments.

\subsubsection{Exploratory studies}
Exploratory studies emphasize evaluating user experience and collecting subjective feedback. They typically involve more open-ended tasks than controlled experiments or online evaluations. 
Exploratory studies \strikethrough{could} can, for example, take place in a controlled lab setting \corey{which is more typical of comparative studies}, yet \strikethrough{be more} \corey{use} open-ended \corey{tasks that} \strikethrough{to} allow music making to occur in a more natural way\corey{.}\strikethrough{instead of comparing two conditions directly, or may take place in less controlled setting.} 
 For example, \citet{bougueng_tchemeube_evaluating_2023} tasked people with exploring their generative AI interface and collected structured questionnaire measures to capture aspects of people's user experience, but do not make strict comparisons between interface designs. 
\strikethrough{Note that a} \corey{Thus,} balance can be struck between controlled and exploratory evaluations, such as giving open-ended tasks in controlled study settings to test generative AI in a way that is closer to real-world music making (e.g., \cite{louie_novice-ai_2020}), or to give structure to data collection in real-world settings (e.g., \cite{ford_reflection_2024}). 

\subsubsection{`In-the-wild' studies}
Research-in-the-wild \cite{benford_performance-led_2013,chamberlain_research_2012} approaches contrast controlled experiments to evaluate generative AI models in their real-world places of use, possibly over extended periods of time. For example, in ethnographic approaches \cite{chamberlain_research_2012,benford_performance-led_2013} the researcher takes observations or field notes, or collects data on patterns of behavior that people have naturally exhibited while making music. For AI music, this type of approach has been used in, e.g.,\strikethrough{the context of} the international AI music song contest \cite{huang_ai_2020}. \alex{The researchers identified how developers and musicians collaborated to create music}	, for example by preferring to curate AI generated content instead of (re-)developing their AI tools. Across the HCI studies on generative music, there are several examples of ethnographic-inspired observations being collected. \citet{fiebrink_toward_2010}, for instance, ``recorded text minutes of [composer's] activities, discussion topics, and specific questions, problem reports, and feature requests'' 
for seminars on their Wekinator \cite{fiebrink_human_2011} system. \citet{bryan-kinns_using_2024} used first-person accounts of music making with a generative AI system over several months to understand how it was appropriated into music making practice.



        

\subsection{User data collection: Quantitative}\label{sec:quant_data}

\alex{The primary quantitative method for evaluating generative AI systems is the use of \emph{questionnaires}}\strikethrough{The key quantitative data collection technique used to evaluate generative AI systems are \emph{questionnaires}}. These are used to quantify both subjective feelings of a system's usability as well as more experiential aspects. Typically, these questionnaires use a Likert-type scale \cite{likert_technique_1932} \alex{to measure user agreement with statements on a scale (e.g., 1–5)}\strikethrough{that rates on a scale, e.g., from 1--5, how much a user agrees or disagrees with a statement}. 



Several standard questionnaire measures exist to evaluate the usability of technology. Common examples are the NASA Task Load Index \cite{hart_nasa-task_2006} and the Standard System Usability Scale \cite{bangor_empirical_2008}. For AI music interfaces, we found several evaluations (e.g., \cite{hunt_empirical_2021}) using the Cognitive Dimensions of Notations questionnaire \cite{blackwell_cognitive_2000}, which assess several cognitive qualities of the interface such as whether the system has many hidden elements of represents information in a diffuse way. 

User experience-oriented scales include the User Experience Questionnaire \cite{laugwitz_construction_2008} and the User Engagement Scale \cite{obrien_practical_2018}. The most influential questionnaire with respect to creative technology within the last 10 years is arguably the Creativity Support Index (CSI) \cite{cherry_quantifying_2014}: a questionnaire designed to test the capacity of a tool to support creativity, offering factors for several important aspects of the creative user experience including: focused attention, enjoyment and collaboration. For AI music, however, we found surprisingly few examples of AI music user studies that have adopted the CSI --- \citet{bougueng_tchemeube_evaluating_2023} \strikethrough{being} a notable exception. We instead observe many examples where researchers have chosen to define their own questionnaires to explore constructs for the AI systems that are not captured in current standardized scales. For instance, \citet{louie_novice-ai_2020} invented their own questions to capture a person's feelings of agency. \citet{ford_towards_2023} identified reflection as an aspect missing from the CSI whilst being an important factor in AI music making \cite{ford_reflection_2024, ford_speculating_2022}. Alongside established measures of creativity support and usability, \citet{bougueng_tchemeube_evaluating_2023} also added questions central to human-AI interaction on trust, perceived authorship, and flexibility. 
\corey{This use of researcher defined questionnaires reflects the dominance of measures of engagement and usability in assessing AI interaction in creativity-related HCI research \cite{cox_reflecting_2025,cox_beyond_2025}.}

Questionnaires are more often used in usability-focused evaluation methodologies such as controlled experiments and less frequently used as part of more experiential evaluations such as in-the-wild studies. HCI researchers also use questionnaires to establish characteristics of their sample under study. For example, the Goldsmith's Musical Sophistication Index \cite{mullensiefen_musicality_2014} offers a standardized metric for musical expertise, helping to identify whether the users of a generative AI system under study have above or below average musical skills.

\subsection{User data collection: Qualitative} 


In addition to quantitative data collection researchers use qualitative data collection to gain greater insight into users' feelings, motivations, and reflections when using \corey{a} generative AI system. It is important to note that triangulation across different data collection approaches (also referred to as mixed methods) is crucial --- using both qualitative and quantitative data can help to demonstrate which features of the user experience are \strikethrough{or are not} improved by AI \corey{or not}, as well as offering insights into why this might be so \cite{bryan-kinns_guide_2023}.

\subsubsection{Interviews}
Interviews are a common technique used in HCI, often to give insights into users' thoughts and feelings on their interaction. They can be structured, semi-structured or fully open-ended \cite{braun_successful_2013}. For generative music user studies, we found that semi-structured and unstructured approaches were common, with workshops or group interviews used for need-finding studies \cite{suh_ai_2021,fiebrink_reflections_2020,louie_novice-ai_2020,frid_music_2020,ford_reflection_2024}. There is no standard set of questions used in interviews for evaluating interaction with generative AI models, nor standard analysis approaches. Results tended to be reported thematically following a process such as Thematic Analysis \cite{braun_reflecting_2019} or using more experimental approaches as in \citet{fiebrink_reflections_2020}, who transcribed their interviews verbatim when reflecting on their extensive experience on AI music. Generative AI studies in music are yet to explore qualitative analysis methods emerging in more modern HCI paradigms \cite{frauenberger_entanglement_2019}, e.g., \cite{scurto_prototyping_2021,rajcic_towards_2024}, which might capture qualities from interviews that Thematic Analysis does not.


\subsubsection{Think-aloud}
Several studies \cite{suh_ai_2021,louie_novice-ai_2020,louie_expressive_2022,frid_music_2020,hunt_empirical_2021} have applied the HCI ``think-aloud'' method to gain insight into how users interact when making music with a generative AI model. In the think-aloud method participants are asked to describe their thought process while performing their task, e.g., while making music with an AI tool \cite{preece_interaction_2015}. Whilst this method can give detailed insight into participant's cognitive process, it can distract users, meaning that certain aspects of the creative user experience such as flow states cannot then be investigated \cite{csikszentmihalyi_flow_1990}. It is also impractical for certain music practices such as live improvisation. An alternative approach is to perform the think-aloud retrospectively \cite{alshammari_when_2015} with participants describing a recording of their composition practice (sometimes referred to as video-cued recall \cite{candy_practice-led_2006})~---~this approach is under-utilized in the literature for generative music.

\subsubsection{First-person perspectives}
The use of methods such as questionnaires and interviews described above \strikethrough{are} \corey{is} borne from a psychology-driven epistemological stance: to identify generalizable models of how people interact with technology. Approaches inspired by Arts and Humanities offer insights into the individuality and subjectivity of how artists have interpreted their use of technology \cite{candy_practice-based_2018}. We identified an increasing trend to report on the use of AI from a first-person perspective \cite{noel-hirst_autoethnographic_2023,ford_reflection_2024,sturm_machine_2019,sturm_generative_2022,ben-tal_how_2021,bryan-kinns_using_2024}, publishing perspectives on how individuals have been able to use and incorporate models into their music-making. In some cases, these are autoethnographies \cite{noel-hirst_autoethnographic_2023,sturm_generative_2022} where a researcher reflects on their own practice by means of capturing data over a long time period. Other examples show collections of first-person accounts \cite{ford_reflection_2024,sturm_machine_2019}.  \citet{sturm_musaicology_2024}'s proposal for a field of AI music studies engages with these methods to explore ways to more meaningfully and critically engage with the broader communities in social sciences and humanities. In contrast to questionnaires, first-person perspectives are more frequently used in user experience-focused evaluations such as in-the-wild studies and less frequently, if at all, in controlled experiments.

\nbk{
        \begin{footnotesize}
            \begin{table}								
\caption{\nbk{Overview of Human-Computer Interaction evaluation approaches.}}\label{tab:hci_evaluation}            								
\centering								
\begin{tabular}{p{.35\linewidth}|p{.1\linewidth}|p{.35\linewidth}|p{.1\linewidth}}								
								
\toprule  								
	\multicolumn{4}{l}{\textbf{Quantitative Methods: Questionnaires}}							\\
		&	\textbf{Acronym}	&	\textbf{Captures}	&	\textbf{Reference}	\\
\hline  	Goldsmith’s Musical Sophistication Index	&	GMSI	&	Musical expertise	&	\cite{mullensiefen_musicality_2014}	\\
\hline  	NASA Task Load Index	&	NASA TLX	&	How complex a task is perceived to be	&	\cite{hart_nasa-task_2006}	\\
\hline  	Standard System Usability Scale	&	SUS	&	The usability of a user interface	&	\cite{bangor_empirical_2008}	\\
\hline  	Cognitive Dimensions of Notations questionnaire	&	CD	&	The usability of a user interface	&	\cite{blackwell_cognitive_2000}	\\
\hline  	User Experience Questionnaire	&	UEQ	&	The usability and experiential aspects of a user interface	&	\cite{laugwitz_construction_2008}	\\
\hline  	User Engagement Scale	&	UES	&	The experiential aspects of a user interface	&	\cite{obrien_practical_2018}	\\
\hline  	Creativity Support Index	&	CSI	&	How well a user interface supports creative work	&	\cite{cherry_quantifying_2014}	\\
\hline  	Reflection in Creative Experience	&	RiCE	&	Types of reflection in creative contexts	&	\cite{ford_towards_2023}	\\
\toprule								
								
	\multicolumn{4}{l}{\textbf{Qualitative Methods}}							\\
		&	\textbf{Examples}	&	\textbf{Captures}	&	\textbf{Reference}	\\
\hline								
	Interviews: Structured, semi-structured, or open-ended	&	\cite{suh_ai_2021,fiebrink_reflections_2020,louie_novice-ai_2020,frid_music_2020,ford_reflection_2024}	&	Insights into users’ thoughts, motivations, and feelings on their interaction	&	\cite{braun_successful_2013}	\\
\hline  	Think-aloud	&	\cite{suh_ai_2021,louie_novice-ai_2020,louie_expressive_2022,frid_music_2020,hunt_empirical_2021}	&	Users' thought processes whilst using a user interface	&	\cite{preece_interaction_2015}	\\
\hline  	Video-cued recall	&	\cite{candy_practice-led_2006}	&	Users' post-hoc thoughts about using a user interface	&	\cite{alshammari_when_2015}	\\
\hline  	Autoethnographies	&	\cite{noel-hirst_autoethnographic_2023,sturm_generative_2022}	&	A researcher's subjective and personal reflections on their musical practice and use of technology	&	\cite{rapp_autoethnography_2018}\\
\hline  	First-person accounts	&	\cite{ford_reflection_2024,sturm_machine_2019}	&	Rich descriptions of users' personal reflections on using AI models	&	\cite{lucero_sample_2019}\\
								
\bottomrule								
\end{tabular}								
\end{table}								
        \end{footnotesize}
}

\subsection{Other HCI evaluation approaches}
There is a wide variety of other evaluation methods used in HCI beyond those we found are most often used when evaluating AI music systems\corey{,} as described in this section. 
Usability metrics such as task completion rates, task time or the number of errors, have been explored to evaluate the usability of a computer music system  \cite{wanderley_evaluation_2002}. However, these are not prominent in generative AI music user studies where music making tends to have no clearly defined goal.
``Wizard of Oz'' was an early popular approach to user studies of human-AI interaction more broadly, where users interact with a user interface whilst a researcher provides feedback through the interface in lieu of an AI. For example, \citet{thelle_how_2022} tested participant's reactions to researchers who performed piano phrases live, acting as an AI  system. Similar examples have been tested with more complex generative music programming languages \cite{bellingham_choosers_2022}.
We also did not find many examples of using physiological measures such as heart rate or eye-tracking to explore people's interaction with music AI tools. In other arts-based HCI studies, physiological measures have been used as proxies of aspects of the user experience such as anxiety or boredom, indicated by participants' heart rate or skin conductance \cite{maier_DeepFlow_2019}. This could be an open area for further research.

\section{Challenges and future work}\label{sec:challenges}
 Despite a large number of previously proposed approaches, a multitude of challenges remain unsolved in the evaluation of generative systems in music. Given the nature of the task, it is unclear if generalizable satisfactory solutions can ever be found for some of these challenges.

\subsection{Validity}\label{sec:validity_and_generalization}
       Although subjective evaluation is often considered the most meaningful way of evaluating the output of generative systems, the results cannot be automatically assumed to be robust or reliable. Subjective evaluation of system output (Sect.~\ref{sec:subjective_eval}) usually relies on survey approaches \cite{hernandez-olivan_subjective_2022}, much as quantitative user data collection relies on questionnaires (Sect.~\ref{sec:quant_data}). However, designing surveys and questionnaires is non-trivial~---~the creation of batteries and psychometrics form  an entire subfield of psychology \cite{gehlbach_measure_2011, sanbonmatsu_impact_2021}. Moreover, \alex{existing questionnaires often fail to capture experiential aspects of human-AI interaction, such as users’ sense of agency}\strikethrough{questionnaires have not yet been developed to cover many experiential aspects of human-AI interaction such as people's feelings of agency}. A considerable number of researchers working on generative music systems lack the background, skill and/or resources to successfully carry out such surveys with valid, reliable, and replicable results \cite{yang_evaluation_2020}. 

        In addition, there is a known bias in people's perception against AI-generated music \cite{shank_AI_2023}. As such, \alex{tests highlighting human vs.\ machine authorship}\strikethrough{any test in which a question about human vs. machine creation is raised} may introduce bias in the results.
        With respect to Turing tests, \citet{hernandez-orallo_twenty_2020} point out that a known validity issue with Turing tests is that the outcomes cannot disambiguate whether the model was a good imitator, or the human was a poor judge.
        
        Given these potential validity issues with subjective evaluation of system output, objective evaluation remains a viable choice to complement listening studies. With objective evaluation, however, there are other validity concerns. 
        \strikethrough{Some of t}\alex{T}hese concerns \alex{often} start with the data and its characteristics: Is the sample size \strikethrough{large enough}\alex{sufficient?} Do the data reflect the targeted homogeneity or heterogeneity? Are there confounding characteristics in the data that complicate drawing conclusions? 
        Another concern is the \alex{validity of the chosen metrics --- do they meaningfully}\strikethrough{how meaningful the used metrics are, i.e., whether the objective metrics properly} represent the evaluation target\alex{, and are observed} \strikethrough{or whether } differences in evaluation results perceptually \alex{significant}\strikethrough{relevant or not.}? Even if some metrics prove to be relevant, individual metrics or criteria \strikethrough{are usually insufficient to provide a broader assessment}\alex{provide sufficient breadth for comprehensive evaluation}; \citet{theis_note_2016} rightly note that ``Good performance with respect to one criterion (...) need not imply good performance with respect to the other criteria.''
        Note that even when targeting very specific criteria (e.g., complexity), subjective impressions might be better predictors of responses than objective measures \cite{hargreaves_experimental_2010}. For that reason, it might make more sense to use subjective impressions of criteria (rated by the listener) as predictors of the overall judgments of aesthetic value \cite{juslin_aesthetic_2023}.
                
        For evaluating user interaction with generative systems using HCI approaches, the main challenge is to balance the ecological validity of the evaluation (how realistic is the study setting) and the generalizability of the results of the study. For example, ``in-the-wild'' studies give in-depth insights into how generative music is used in real-world places of music making such as performances on stage or music making at home, but the findings are tied to the study's cultural context and individual musicians making it hard to generalize from the results. Certain protocols such as ``think-aloud'' also affect ecological validity because subjects tend not to speak aloud about their thought processes when making music. Likewise, the ``Wizard of Oz'' protocol has poor ecological validity as understanding how people interact with pretend AI tools can be different to how these systems work once actually deployed.
        
        Thus, both internal and external validity remain core challenges of evaluating generative music systems. This is true for both subjective and objective approaches. 

\subsection{Perceptual and musical relevance of objective metrics}
    Revisiting the objective metrics introduced in Sect.~\ref{sec:output}, it can be observed that the most common metrics compare training data characteristics with characteristics of the generated data in one way or the other. A major differentiation between such metrics is the space and the dimensions in which different metrics approach such a comparison. 
    On the one hand, learned embeddings such as VGGish \cite{hershey_cnn_2017} and a variety of other embeddings have been utilized for the FAD or related metrics, on the other hand there are custom-designed low-level statistical descriptors such as pitch range, pitch class histograms, etc.~\cite{yang_evaluation_2020}. This has considerable impact on the interpretability of the descriptors; while a learned descriptor such as VGGish cannot be interpreted directly, custom-designed descriptors tend to be more interpretable. However, high interpretability does neither mean that the descriptor is relevant for assessment nor that it is perceptually or musically meaningful.

    Perceptual studies of specific descriptors are necessary to understand their relevance and meaning. Simply finding  a difference between two set of data with respect to one descriptor does not automatically mean that these data are different from a perceptual point of view. \alex{Recent studies on the suitability of learned embeddings for evaluation seem to focus on overall relevance for summary aesthetic judgments without considering interpretability or musical meaning \cite{huang_aligning_2025, grotschla_benchmarking_2025}.}

    Furthermore, even descriptors known for their perceptual validity can be more or less meaningful depending on context and scenario. 
    For instance, it has been demonstrated that people are incredibly sensitive to the statistical distribution of pitches in a piece of music, and can even learn new musical systems and grammars based on such statistical inference (e.g., \cite{krumhansl_tonality_2000, huron_sweet_2006, temperley_cognition_2004, loui_imaginings_2023}).
    However, in general, not all musical representations are equal, and this can have a sizeable impact in the perceptual relevance of any given feature or metric. 
    For instance, most music is tonal meaning that it is (at least temporarily) in a given key or mode and has a stable tonic.
    For such tonal music, listeners are very sensitive to notes that are outside of the key (i.e., 'wrong notes') \cite{krumhansl_tonality_2000, dowling_pitch_1991, raman_effects_2016}.
    As such, measuring the statistical distribution of pitches in relation to that tonic (as scale degrees or musical intervals from a tonic) carries a different musical and perceptual relevance compared to the distribution of all pitch classes measured in a tonic-agnostic way.
    Thus, it is not only necessary to validate whether certain descriptors have perceptual relevance per se, but also in what (e.g., tonal or stylistic) context they are extracted.
    But even given a set of relevant descriptors we can only guess how exhaustive this set is. At the very least, the number and type of relevant descriptors is genre-dependent, and is quite possibly indefinite.

\subsection{Reproducibility}
    Reproducibility has long been identified as a problem in the machine learning community \cite{sonnenburg_need_2007}. For software-based technologies and approaches, the pure description of research in a paper is increasingly considered insufficient and the publication of well-documented open-source code has been identified as one important part of a solution \cite{vandewalle_reproducible_2009, mcfee_open-source_2019}. In addition, as most machine learning is data-driven, understanding the training data and the test data is crucial. Unlike other systems for other machine learning tasks, generative systems often use massive amounts of unlabeled (and potentially unpublished) data of potentially unclear origin and with unclear data curation approaches, meaning that the training of such systems cannot be reproduced by unaffiliated parties.  
    
    For generative systems, we introduce the following levels of reproducibility with increasing level of transparency:
    \begin{inparaenum}[(i)]
        \item \textit{publication of an academic text}, describing the method and approaches,
        \item \textit{publication of all raw results}, including the generated music in order to reproduce the result-based conclusions, 
        \item \textit{publication of the generative system} itself (e.g., through an API) to allow reproducing the results with a clear documentation of the system prompts and parameter settings from the study,
        \item \textit{publication of documented source code} of the pre-trained system in order to allow in-depth understanding of architectural details and parameters not published otherwise,
        \item \textit{publication of training source code} for the generative system to share details on data processing and training methodologies,
        \item \textit{publication of training data statistics} to improve transparency around data distribution and characteristics, possible bias, and other details,
        \item \textit{publication of training data} and source code for data pre-processing and curation, and
        \item \textit{publication of data acquisition and curation strategies} to be transparent about potential bias, data licenses, and fair data use.
    \end{inparaenum}
    While we acknowledge that constraints exist that do not always allow for publication of every single detail, we call for full transparency as the goal of any scientific work in this area to the extent possible.

    In addition, HCI studies of generative systems require the publication of user data collected such as questionnaire results, interview transcripts, music generated, and recordings of human interaction with generative system if the studies are themselves to be reproducible by other researchers. There are substantial privacy and practical challenges to making such data available and shareable, not least the lack of any standards for sharing user study data in HCI to date.


\subsection{Need for de-facto evaluation standards}  

    As described above, there is currently a multitude of evaluation methodologies and metrics across studies in the field that prevents results from being comparable to each other. This means that the capabilities and shortcomings of systems are not consistently assessed, and no meaningful conclusions regarding the progress of the field can be drawn. Clearly, there is a need for de-facto evaluation standards \alex{(compare also \citet{xiong_comprehensive_2023, zhou_analysis_2023}). More specifically, standards are needed} with respect to
    \begin{inparaenum}[(i)]
        \item assessment targets,
        \item evaluation methodology,
        \item commonly used, publicly available reference (test) data sets, and
        \item agreement on a base set of evaluation metrics that allows for future extension with additional metrics to avoid metric overfitting. Even an \strikethrough{widely used} imperfect set of metrics can help a field moving forward, as the continued use of BSSEval metrics SDR, SAR, and SIR~\cite{vincent_performance_2006} for source separation systems shows~---~despite widely known shortcomings \cite{emiya_subjective_2011, fox_modeling_2007, gupta_perceptual_2015}.
    \end{inparaenum}    
%
These metrics might complement HCI approaches where there are no de-facto standards for evaluation in general. Whilst ``mixed-methods'' is a current methodological trend, HCI studies are designed to respond to the goals of the evaluation, drawing from both quantitative and qualitative approaches. As such, de-facto standards and benchmarks might not be as meaningful in HCI evaluations where it is more important to understand the features of the evaluation technique used and the user data collected.


\ifthenelse{\equal{\bWithRAI}{true}}
{

\subsection{Responsible AI}  

Responsible AI (RAI) builds on Responsible Innovation's \cite{stilgoe_developing_2013} concerns of ensuring that ``the processes and outcomes of research are aligned with societal values'' \cite{jirotka_responsible_2017}, and argues for developing AI which is accountable, ethical, fair, and explainable \cite{ozmen_garibay_six_2023}. RAI concerns for AI music include tracing how and where data used in generative models have been sourced, as well as global challenges such as their environmental impact. Evaluating RAI is an underexplored area, especially with generative art systems \cite{piskopani_responsible_2023} such as music. 

    \subsubsection{Explainability}
        The explainability and interpretability of a system's output is a major contributor to building trust with the user and AI accountability. As systems get more complex and less understandable, critical questions arise with respect to usage and deployment of these systems \cite{kambhampati_changing_2022}, including in the arts \cite{bryan-kinns_reflections_2024}.
        
        Explainability has been framed as based on four principles \cite{phillips_four_2021}, namely that the system
        \begin{inparaenum}[(i)]
            \item delivers ``accompanying evidence or reasons for outcomes and processes,'' 
            \item provides ``explanations that are understandable to individual users,'' 
            \item provides ``explanations that correctly reflect the system's process for generating the output,'' and 
            \item ``only operates under conditions for which it was designed and when it reaches sufficient confidence in its output.''
        \end{inparaenum}
        However, \citet{vilone_notions_2021} find that there is no general consensus on how such explanations can be defined and how their validity can be assessed. Moreover, these forms of explanation may not be appropriate in more aesthetic and subjective fields such as music and the arts \cite{bryan-kinns_exploring_2024}.

        \citet{barredo_arrieta_explainable_2020} point out that the origin of the explanation can be either through a transparent system or post-hoc through analyzing the model output. Thus, \citet{Batlle-roca_transparency_2023} find that a comparison between evaluation strategies is not possible due to a ``lack of formality and a clear definition on an explanation task.'' This lack of evaluation strategies for explainable AI, and the subjective and experiential nature of the arts, poses significant challenges for defining rigorous evaluation metrics for the explainability of generative music systems.


    \subsubsection{Bias}
        Machine learning models have been shown to exhibit various kinds of bias. The root cause of bias tends to relate to the distribution of the training data emphasizing the importance of properly and transparently curating and documenting training data \cite{bender_dangers_2021} (a complete assessment of bias in machine learning should include a detailed analysis of the training data and their curation). 
         For example, models from natural language processing have shown undesirable biases towards gender \cite{basta_evaluating_2019, kurita_measuring_2019}, race \cite{tan_assessing_2019}, and mentions of disability \cite{hutchinson_social_2020}. Computer vision models have also been shown to be biased even with balanced training data \cite{wang_balanced_2019}, for instance, showing bias towards higher income settings \cite{de_vries_does_2019} and amplifying gender bias \cite{wang_balanced_2019}.


        Bias in music datasets is a known problem. \citet{ehmann_music_2011}, for instance, describe bias towards western popular music in datasets for structural segmentation as early as 2011. This bias can be observed in the output of modern systems as well \cite{barnett_ethical_2023, bryan-kinns_exploring_2024}. Biases in music training datasets for generative systems might lead, for example, to marginalization of non-mainstream musical styles \cite{Bryan-Kinns_explainable_2024} or gender or style imbalance of generated content. The assessment of bias in machine learning is, however, a multi-faceted problem without a single, clear solution \cite{mehrabi_survey_2021}, and requires an informed, targeted approach.


    \subsubsection{Ethical use of data}\label{sec:ethical_data}
        Given the data-driven nature of modern machine learning systems, it can be argued that ethical data acquisition approaches, guiding principles, and transparency on data content and curation are crucial for the holistic evaluation of a machine learning system \cite{jo_lessons_2020}. Besides the aforementioned threat of bias, one of the biggest concerns regarding the training data for generative music systems is whether the training data are sourced with the content creators' consent and without violating copyright or licensing terms \cite{barnett_ethical_2023, epstein_art_2023} and whether it meets creators' ethical restrictions on use  \cite{benjamin_towards_2019}.
        Without transparency around training data, there is a risk of datasets ``hidden and guarded under the guise of proprietary assets'' \cite{birhane_large_2021}. Thus, \citet{birhane_large_2021} propose the dissemination of ``dataset audit cards,'' listing goals, curation procedures, and known shortcomings. Similarly, \citet{srinivasan_artsheets_2021} propose a set of questions to compile a datasheet for datasets \cite{gebru_datasheets_2021} containing artworks.
        There also exist industry attempts to certify fair training data use;\footnote{\url{https://www.fairlytrained.org/}, last access date Jun 25, 2024} the trust in and impact of such efforts will need to be assessed in the future.
        

    \subsubsection{Energy consumption and resource use}
        With the ever-increasing model complexity of generative models come increasing requirements for energy and other resources \cite{strubell_energy_2019, chien_reducing_2023, berthelot_estimating_2024}. Large models have a very high energy consumption during training and non-negligible energy consumption during inference, raising concerns about environmental impact. \citet{bender_dangers_2021} also note that excessive use of resources might unfairly impact marginalized communities, which are more likely to be harmed by negative environmental consequences. 
        
        Adding some form of assessment of environmental impact to the evaluation strategy could raise broader awareness of these issues and encourage more responsible use of AI as well as research on more energy-efficient forms of machine learning models and corresponding hardware. 
        \citet{henderson_towards_2020}, for instance, propose a framework to track energy consumption and carbon emissions to promote energy-responsible research. In the context of generative audio models, \citet{douwes_is_2023} propose to measure energy consumption of both training and inference in relation to the subjective output quality of music generation and analysis systems (see also \cite{douwes_energy_2021}).
        
}{}
              
\section{Conclusion}\label{sec:conclusion}
We presented an overview of the state-of-the-art in evaluating generative systems in music from the perspective of both the system output and the usability of the system. We categorized different evaluation goals and targets, as well as corresponding methodologies and metrics, and concluded that the current state of system assessment makes it difficult to 
\begin{inparaitem}[]
    \item generalize results,
    \item compare state of the art systems objectively, and
    \item measure progress in the field.
\end{inparaitem}
The main challenges identified include
\begin{inparaenum}[(i)]
    \item the perceptual and musical meaningfulness of current evaluation metrics,
    \item the internal and external validity of common experimental setups, and
    \item the lack of reproducibility,
\end{inparaenum}
emphasizing the need for de-facto evaluation standards adopted by the research community. While the direct assessment of aesthetics  might ``remain out of reach in our lifetime and perhaps forever'' \cite{galanter_computational_2012}, there is a need for methodologies and metrics that give us at least a glimpse into some aspects of quality.

\ifthenelse{\equal{\bWithRAI}{false}}
{
In addition to the evaluation targets presented above, there exist other evaluation targets of interest. Of note could be targets that can be summarized under the umbrella of Responsible AI \cite{piskopani_responsible_2023}. These include, e.g., 
\begin{inparaenum}[(i)]
	\item \textit{explainability}: the increasing complexity of machine learning systems puts forth questions with respect to usage and deployment of these systems \cite{kambhampati_changing_2022}, including in the arts \cite{bryan-kinns_reflections_2024}, however, standardized evaluation strategies do not exist \cite{Batlle-roca_transparency_2023};
	\item \textit{bias}: bias of machine learning systems is a known problem \cite{barnett_ethical_2023, bryan-kinns_exploring_2024} and can~---in the case of generative music systems---~lead to marginalization of non-mainstream musical styles \cite{Bryan-Kinns_explainable_2024};
	\item \textit{ethical use of data}: ethical data acquisition, guiding principles, and transparency on data content and curation are crucial for the holistic evaluation of a machine learning system \cite{jo_lessons_2020}, as the call for ``dataset audit cards'' \cite{birhane_large_2021} and formalized datasheets for datasets \cite{srinivasan_artsheets_2021, gebru_datasheets_2021} emphasize;
	\item \textit{resource use} \cite{strubell_energy_2019, chien_reducing_2023, berthelot_estimating_2024}: the high energy consumption of today's models can be linked to environmental impacts; the reporting of carbon emissions \cite{henderson_towards_2020} or the relation of energy consumption to the subjective output quality of generative music systems \cite{douwes_is_2023} could promote energy-responsible research.
\end{inparaenum}}{}

Given the complexity and open-endedness of the task, one should not forget about other ways of assessing or engaging in a dialogue with generative systems. Musicology has a long history of engaging critically with new pieces and forms of music, and traditional modes of assessment should not be discarded as invalid approaches to analyzing and evaluating music, although~---as \citet{sturm_musaicology_2024} point out---~the large scale generation of music creates new challenges for these approaches.
Artistic inquiry is another form of assessing system output that can create societal awareness. For instance, artists have a  history of exposing bias and discrimination in generative AI systems \cite{gaskins_interrogating_2022, small_black_2023}. These artistic discourses offer a lens through which to explore future values and metrics of evaluation beyond the state of the art surveyed in this paper. 

Furthermore, it became clear in writing this paper that for addressing these challenges interdisciplinarity is a necessity, as an exhaustive system evaluation requires expertise not only in the field of
\begin{inparaitem}[]
    \item   machine learning, but also in
    \item   music theory and musicology,
    \item   psychology, 
    \item   human computer interaction, and possibly others.
\end{inparaitem} 
In our view, this is especially true for generative systems for music as music is a fundamental form of human creativity, social interaction, and intangible cultural heritage which itself has defied evaluation for millennia.

\begin{acks}
\corey{Corey Ford was a PhD student at Queen Mary University of London's UKRI Centre for Doctoral Training in AI and Music at the inception of this work (grant number EP/S022694/1).}
\end{acks}

\bibliographystyle{ACM-Reference-Format}
\bibliography{2024-Survey-GenEval, 2025_alex}

\end{document}